\documentclass[a4paper,fleqn,usenatbib,useAMS]{mnras}

\usepackage[dvipsnames]{xcolor}
\usepackage{tikz,hyperref}

\definecolor{lime}{HTML}{A6CE39}
\DeclareRobustCommand{\orcidicon}{
	\begin{tikzpicture}
	\draw[lime, fill=lime] (0,0) 
	circle [radius=0.13] 
	node[white] {{\fontfamily{qag}\selectfont \tiny ID}};
	\draw[white, fill=white] (-0.0625,0.095) 
	circle [radius=0.007];
	\end{tikzpicture}
	\hspace{-2mm}
}

\foreach \x in {A, ..., Z}{\expandafter\xdef\csname orcid\x\endcsname{\noexpand\href{https://orcid.org/\csname orcidauthor\x\endcsname}
			{\noexpand\orcidicon}}
}

\usepackage{graphicx}	
\usepackage{amsmath}	
\usepackage{amssymb}	
\usepackage{multicol}        
\usepackage{bm}		
\usepackage{pdflscape}	
\usepackage[encapsulated]{CJK}
\usepackage{ucs}
\usepackage[utf8x]{inputenc}
\usepackage{multirow}
\usepackage{booktabs,caption}
\usepackage[flushleft]{threeparttable}
\usepackage{adjustbox}
\usepackage{rotating} 
\usepackage{lscape}

\newcommand{\msol}{M$_{\odot}$}

\newcommand{\ha}{H$\alpha$}
\newcommand{\hb}{H$\beta$}
\newcommand{\hg}{H$\gamma$}
\newcommand{\hd}{H$\delta$}

\newcommand{\heiiwr}{He{\,\scshape ii}\,$\lambda4686$}

\newcommand{\civwrrb}{C{\,\scshape iv}\,$\lambda\lambda5806$}

\newcommand{\ergcms}{erg\,cm$^{-2}$\,s$^{-1}$}

\newcommand{\hei}{He{\,\scshape i}}
\newcommand{\heii}{He{\,\scshape ii}}
\newcommand{\cii}{C{\,\scshape ii}}
\newcommand{\ciii}{C{\,\scshape iii}}
\newcommand{\civ}{C{\,\scshape iv}}
\newcommand{\cv}{C{\,\scshape v}}
\newcommand{\cvi}{C{\,\scshape vi}}
\newcommand{\nia}{[N{\,\scshape i}]}
\newcommand{\nii}{[N{\,\scshape ii}]}

\newcommand{\nv}{N{\,\scshape v}}
\newcommand{\ovi}{O{\,\scshape vi}}

\newcommand{\ov}{O{\,\scshape v}}
\newcommand{\oiii}{[O{\,\scshape iii}]}
\newcommand{\oii}{[O{\,\scshape ii}]}
\newcommand{\oi}{[O{\,\scshape i}]}

\newcommand{\sii}{[S{\,\scshape ii}]}
\newcommand{\siii}{[S{\,\scshape iii}]}

\newcommand{\cliii}{[Cl{\,\scshape iii}]}

\newcommand{\ariv}{[Ar{\,\scshape iv}]}
\newcommand{\arv}{[Ar{\,\scshape v}]}

\newcommand{\neiii}{[Ne{\,\scshape iii}]}
\newcommand{\nevii}{Ne{\,\scshape vii}}
\newcommand{\neviii}{Ne{\,\scshape viii}}
\newcommand{\oviii}{O{\,\scshape viii}}
\newcommand{\ovii}{O{\,\scshape vii}}
\newcommand{\oiia}{[O{\,\scshape ii]}\,$\lambda3727$}

\usepackage[T1]{fontenc}
\usepackage{ae,aecompl}

\usepackage{newtxtext,newtxmath}

\title[PNe with WR CSPN -- III]
{Planetary nebulae with Wolf-Rayet-type central stars - III. A detailed view of NGC\,6905 and its central star}
\author[G\'{o}mez-Gonz\'{a}lez et al.]{V.~M.~A.\,G\'{o}mez-Gonz\'{a}lez\thanks{E-mail:\,mau.gglez@gmail.com, vmagg@astro.physik.uni-potsdam.de}$^{1,6\orcidA{}}$, G.\,Rubio$^{2,3\orcidB{}}$, J.~A.\,Toal\'{a}$^{1\orcidC{}}$, M.~A.\,Guerrero$^{4\orcidD{}}$, L.\,Sabin$^{5\orcidE{}}$,  H.\,Todt$^{6}$, \newauthor{V. G\'omez-Llanos$^{5\orcidF{}}$, G.\,Ramos-Larios$^{2,3\orcidH{}}$ and Y.~D.\,Mayya$^{7\orcidI{}}$}
\\
$^{1}$Instituto de Radioastronom\'{i}a y Astrof\'{i}sica, UNAM Campus Morelia, Apartado postal 3-72, 58090 Morelia, Michoac\'{a}n, Mexico\\
$^{2}$Instituto de Astronom\'{i}a y Meteorolog\'{i}a, CUCEI, Universidad de Guadalajara, Av. Vallarta 2602, Arcos Vallarta, 44130 Guadalajara, Mexico\\
$^{3}$CUCEI, Universidad de Guadalajara, Blvd. Marcelino Garc\'\i a Barrag\'an 1421, 44430, Guadalajara, Jalisco, Mexico \\
$^{4}$Instituto de Astrof\'{i}sica de Andaluc\'{i}a (IAA-CSIC), Glorieta de la Astronom\'{i}a S/N, 18008 Granada, Spain\\
$^{5}$Instituto de Astronom\'{\i}a, Universidad Nacional Aut\'onoma de M\'exico, Apdo. Postal 106, 22800 Ensenada, B.C., Mexico\\
$^{6}$Institute for Physics and Astronomy, Universit\"{a}t Potsdam, Karl-Liebknecht-Str. 24/25, D-14476 Potsdam, Germany\\
$^{7}$Instituto Nacional de Astrof\'{i}sica, \'{O}ptica y Electr\'onica, Luis Enrique Erro 1, Tonantzintla 72840, Puebla, Mexico\\ 
}

\pubyear{2021}

\begin{document}
\label{firstpage}
\pagerange{\pageref{firstpage}--\pageref{lastpage}}
\maketitle

\begin{abstract}
We present a multi-wavelength characterisation of the planetary nebula (PN) NGC\,6905 and its [Wolf-Rayet]-type ([WR]) central star (CSPN) HD\,193949. Our Nordic Optical Telescope (NOT) Alhambra Faint Object Spectrograph and Camera (ALFOSC) spectra and images unveil in unprecedented detail the high-ionization structure of NGC\,6905. The high-quality spectra of HD\,193949 allowed us to detect more than 20 WR features including the characteristic O-bump, blue bump and red bump, which suggests a spectral type no later than a [WO2]-subtype. Moreover we detect the \nevii\ and \neviii\ broad emission lines,
rendering HD\,193949 yet another CSPN with $T_\mathrm{eff}\lesssim150$~kK exhibiting such stellar emission lines. We studied the physical properties ($T_\mathrm{e}$ and $n_\mathrm{e}$) and chemical abundances of different regions within NGC\,6905 including its low-ionization clumps;  abundances are found to be homogeneous.
We used the PoWR stellar atmosphere code to model the spectrum of HD\,193949, which is afterwards used in a photoionization model performed with {\scshape Cloudy} that reproduces the nebular and dust properties for a total mass in the 0.31--0.47~M$_{\odot}$ range and a mass of C-rich dust of $\sim$2 $\times10^{-3}$~M$_{\odot}$. Adopting a current stellar mass of 0.6~M$_{\odot}$, our model suggests an initial mass $\sim$1~M$_\odot$ for HD\,193949, consistent with the observations.
\end{abstract}





\section{INTRODUCTION}

By the end of their lives, low- and intermediate-mass stars 
(0.8~M$_{\odot} \lesssim M_\mathrm{ZAMS} \lesssim 8$~M$_{\odot}$) 
produce copious ejections of material when evolving through the asymptotic giant branch (AGB) phase before the final white dwarf (WD) stage. 
Up to half of the initial mass of the star is deposited into the interstellar medium (ISM) through a slow and dense wind \citep[$\text{v}_\mathrm{AGB}=10-30$~km~s$^{-1}$, 
$\dot{M}\approx10^{-6}-10^{-5}$~M$_{\odot}$~yr$^{-1}$;][]{VW1993}. 
Owing to the low effective temperature in this phase ($T_\mathrm{eff} < 10^{4}$~K), a dusty envelope is formed \citep[see][]{Mauron2013}.
After ejecting its outer layers, the star evolves 
into the post-AGB phase increasing its $T_\mathrm{eff}$ and subsequently developing a
strong ionizing UV flux and a fast stellar wind 
\citep[$\text{v}_{\infty}=500-4000$~km~s$^{-1}$; e.g.,][]{Guerrero2013}.
The combination of these
two effects sweeps, compresses, heats and ionizes the previously ejected AGB material, forming
a planetary nebula \citep[PN;][]{Kwok2000}.

As a result of the vast number of works in the past decades, increasing evidence suggest that binaries could play a key r\^{o}le in PN shaping: producing apparent spherical morphologies \citep[see][]{Guerrero2020} 
as well as extremely collimated structures \citep[see][]{Frank2018}.
Changes in the orbital parameters caused by the evolution of the primary star affect the mass-loss \citep{Iaconi2017}.
Moreover, binarity can produce changes in stellar structure and surface abundance,
creating H-poor stars \citep{Lau2011} and reflecting in the PN chemical composition \citep{Wesson2018}.

\begin{figure*}
\begin{center}
\includegraphics[width=\linewidth]{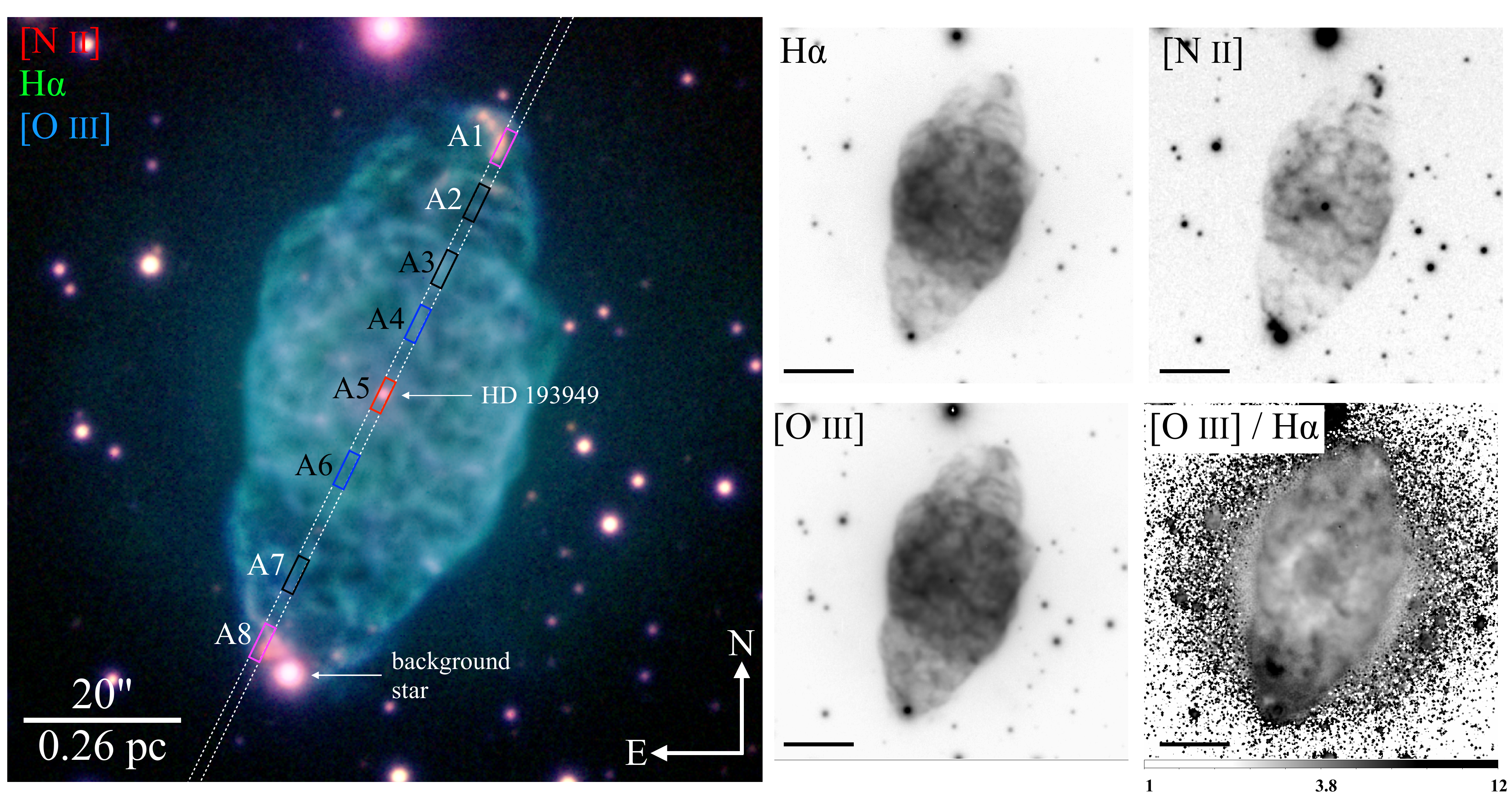}
\caption{({\itshape left}) NOT ALFOSC colour-composite image of NGC\,6905
(\oiii\ -- blue, \ha\ -- green, \nii -- red).
The white dashed-line region represents the NOT ALFOSC slit at PA of 155$^\circ$.
The solid-line rectangles (0.75$\times$5~arcsec), labeled from A1 to A8, indicate our spectra extraction regions.
The position of the CSPN (HD\,193949)
and a background star are indicated with arrows.
({\itshape right}) Individual \ha, \nii\ and \oiii\ intensity images and line intensity \oiii/\ha\ ratio map. Each panel have the same FoV.} 
\label{fig:ngc6909_opt}
\end{center}
\end{figure*}

About 10 per cent of the central stars of planetary nebulae (CSPNe) are H-poor 
with strong and broad emission lines from He, C, N and O in their spectra.
These spectral features resemble those typical in massive stars in their Wolf-Rayet (WR) phase
\citep[see][]{2007Crowther} and can be classified following the same scheme 
\citep[see][]{Crowther1998,Acker2003} as C-rich ([WC]), N-rich ([WN]) and with strong O emission lines ([WO])\footnote{The square brackets ([WR]) are used to denote that they are low-mass stars instead of massive WR stars  (Population I).}, 
with sub-classifications assigned depending on the ratio of the line strengths in their spectra.
The most numerous [WR]-type stars are those of the carbon sequence, followed by the [WO]-types \citep[see table~2 in][and references therein]{2020Weidmann}, with a few cases of 
[WN]-types and transitional [WN/C]-types \citep[e.g.,][]{Todt2010,Todt2013,Miszsalski2012}.

Statistical analyses of PNe with [WR]-type CSPNe (hereinafter WRPN) have
unveiled differences when compared to PNe with H-rich CSPNe.
High-resolution spectroscopic observations have shown that WRPNe exhibit larger 
expansion velocities and present a higher degree of turbulence than non-WRPNe,
very likely due to their higher wind mechanical luminosities or the presence of 
highly-collimated outflows \citep[][]{Gesicki2003, Medina2006,Rechy2017}.
WRPNe have also been found to have higher N and C abundances, suggesting that they are
formed as a result of $\sim$4~M$_{\odot}$ stars \citep{GR2013}.
It has also been suggested that there is a relation between the decrease in
electron density with decreasing spectral subtype as a sign
of evolutionary sequence from late to early [WC]-types \citep[][]{Acker1996,Gorny2000}.
However, the N/O abundance ratio versus spectral type has shown that stars with different
masses can evolve through the same [WC] state \citep[][]{Pena1998,Pena2001}.

In \citet[][hereinafter Paper~I]{2020Gomez}, we started a series of works dedicated 
to a comprehensive characterisation of WRPNe.
In Paper~I we noted that statistical studies of WRPNe are
helpful unveiling their properties as a population, but these do not provide the details that particular
studies can offer us when combining multi-wavelength observations with theoretical predictions.
Our analysis of the high-excitation WRPN NGC\,2371 and its [WO1] CSPN  
WD\,0722$+$295 presented in Paper~I helped us to unveil its physical properties, origin and kinematical structure.

In this paper, we present a comprehensive multi-wavelength characterisation of NGC\,6905 (PN G\,061.4-09.5), a.k.a. the "Blue Flash Nebula" for its characteristic colours. NGC\,6905 is a high-excitation WRPN with a clearly clumpy morphology (see Fig.~\ref{fig:ngc6909_opt}), located at a distance of 2.7~kpc \citep{Gaia2020}.
It is composed of a central roundish cavity with angular radius $\sim$19~arcsec (0.25~pc)
and a pair of extended V-shaped structures extending towards the NW and
SE directions with a position angle (PA) of $\sim$160$^{\circ}$
and with a total extension $\sim$80~arcsec (1~pc).
These bipolar structures harbour a pair of low-ionization knots at their tips. 
These low-ionization features expand faster than the central regions of NGC\,6905 with an inclination
of 60$^\circ$ with respect of the plane of the sky \citep{Sabbadin1982} and have been
suggested to be responsible for giving this WRPN its bipolar morphology \citep{Cuesta1993}.

The CSPN of NGC\,6905 has been classified as [WO]-type star.
Due to the significant \ovi\ emission lines, it has been classified as a [WO2]-subtype \citep[][]{Acker2003}.
Spectroscopic studies in UV of NGC\,6905 such as those presented by \citet{Keller2011,2014Keller} have allowed stellar atmosphere modeling of HD\,193949 implying a $T_\mathrm{eff}$ in the range of 150--165~kK. These authors also found Ne to be overabundant with a mass fraction of 0.02.

This paper is organised as follows.
In Section~2, we describe the observations used in this work.
In Section~3, we present the analysis of the spectra and images.
A discussion is presented in Section~4.
Finally, our summary and conclusions are listed in Section~5.

\section{OBSERVATIONS AND DATA PREPARATION}

\subsection{Nordic Optical Telescope}

Optical images and spectra of NGC\,6905 were obtained with
the Alhambra Faint Object Spectrograph and Camera 
(ALFOSC)\footnote{\url{http://www.not.iac.es/instruments/alfosc/}} at the 2.5~m Nordic Optical Telescope (NOT) of the 
Observatorio del Roque de los Muchachos (ORM) in La Palma, Spain.

The images were obtained on 2009 July 21, using the
\ha\ ($\lambda_\mathrm{c}$=6563~\AA\ and $\Delta \lambda$=33~\AA), 
\nii\ ($\lambda_\mathrm{c}$=6584~\AA\ and $\Delta \lambda$=36~\AA)
and \oiii\ ($\lambda_\mathrm{c}$=5010~\AA\ and $\Delta \lambda$=43~\AA) 
narrow-band filters. Two images of 450~s were obtained in each filter.
The images were processed using standard {\scshape iraf} routines \citep{Tody1993}
and are presented in Figure~\ref{fig:ngc6909_opt}, together with a colour-composite picture
and a \oiii/\ha\ ratio map.

NOT spectroscopic observations were carried out on 2020 July 26.
The EEV 231–42 2K$\times$2K CCD was used, providing a
pixel scale of 0.211~arcsec pixel$^{-1}$ and a field of view (FoV) of 7~arcmin.
A spectrum at a position angle (PA) of 155$^\circ$ was obtained
using the grism \#7, which covered the spectral range
of 3650--7110~\AA\ and
a dispersion of $\sim$1.7\,\AA\,pixel$^{-1}$.
The total observing time was of 2400~s, which was split into two exposures of 1200~s.
We adopted a slitwidth of 0.75~arcsec, with spectral resolution of $\sim$5~\AA.
The seeing during the observations was $\sim$1~arcsec.
The details of the spectroscopic observations are presented in Table\ref{tab:obs}.

The spectra were also reduced and analyzed using standard {\scshape iraf} routines, which include bias subtraction and flat-fielding. The wavelength calibration was performed using HeNe arc lamps and the flux calibration by using the standard star Wolf 1346.

\begin{figure*}
\begin{center}
\includegraphics[angle=0,width=1.0\linewidth]{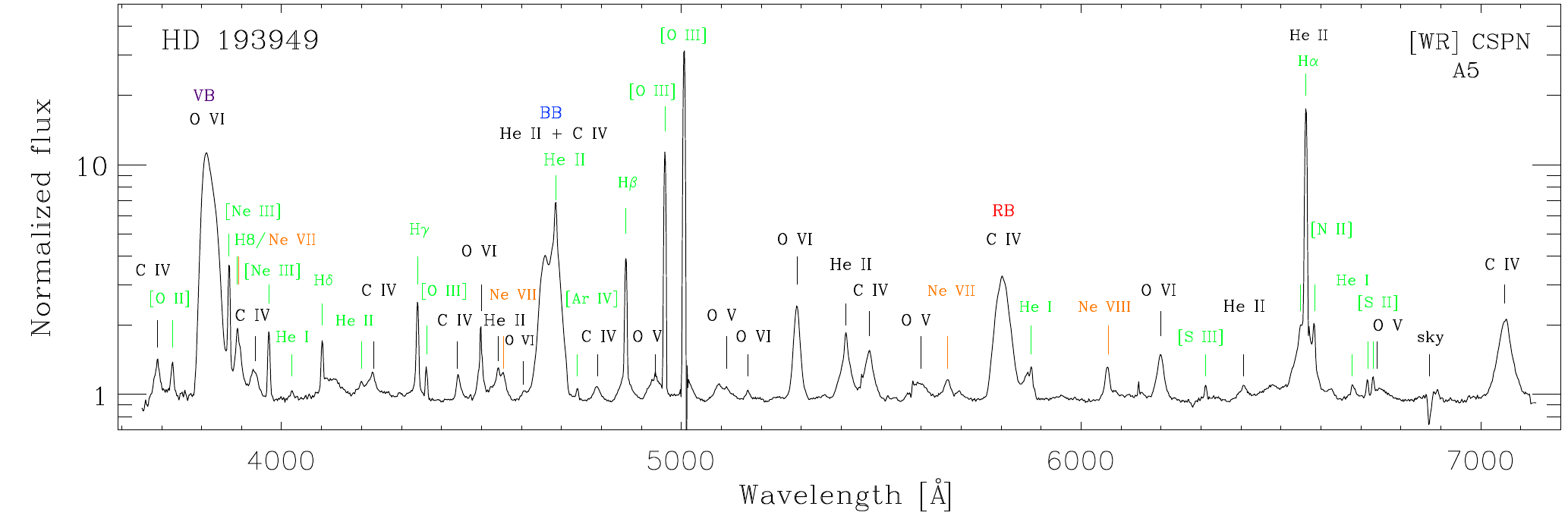}
\caption{NOT ALFOSC spectrum of HD\,193949, the CSPN of
  NGC\,6905, extracted from region A5 defined in Fig.~\ref{fig:ngc6909_opt} (left panel). 
  The most prominent broad WR features like the violet bump (VB) at $\sim\lambda3820$~\AA,
  the blue bump (BB) 
  at $\sim\lambda4686$~\AA\ and the red bump (RB) at $\sim\lambda5806$~\AA\ are marked with their respective colours, whilst other WR features are indicated in black.
  Broad Ne lines of high ionization potential are indicated with 
  orange labels, and the narrow nebular lines are labeled in green.
  The spectrum is shown normalized to the best-fit continuum spectrum.
  See table Table~\ref{tab:elines} for details.}
\label{fig:spec_g7}
\end{center}
\end{figure*}

In order to study the ionization structure of NGC\,6905, we have extracted spectra from different regions defined along the slit using the {\scshape iraf} task \textit{apall}. We first extracted the 1D spectrum of the CSPN tracing the stellar continuum along the 2D spectrum. This information was then used as a reference to trace the nebular spectra along the 2D spectra, as these regions do not show continuum emission. The location of the extracted regions, labelled A1--A8, are shown in Figure~\ref{fig:ngc6909_opt}, all of them with equal sizes of 0.75$\times$5~arcsec. The corresponding extraction region for HD\,193619 is A5.
Finally, we note that all extracted spectra were corrected for extinction by using the $c$(H$\beta$) value estimated from the Balmer decrement method with an intrinsic H$\alpha$/H$\beta$, H$\gamma$/H$\beta$ and H$\delta$/H$\beta$ ratios of 2.86, 0.45 and 0.25, respectively, which corresponds to mean values for $T_{\rm e}$ in the range 2000 K and 12,000 K and $n_{\rm e}$ between 100~cm$^{-3}$ and 1200~cm$^{-3}$ \citep[see][]{Osterbrock2006} and using the reddening curve of \cite{Cardelli1989} with $R_\mathrm{V}$=3.1. The spectrum of HD\,193949 is presented in Figure~\ref{fig:spec_g7} and the spectra from different regions will be discussed in Section~\ref{sec:physical}.

\begin{table}
\small\addtolength{\tabcolsep}{-3pt}
\begin{center}
\caption{Spectroscopic observing log.}
\begin{tabular}{ccccccc}
\hline
Grating& Type   &Name&\multicolumn{2}{c}{Coordinates (J2000)}& Exposure      & Airmass\\
       &        &          & R.A.       & Dec.               &    time (s)   &    \\
\hline
grism7 &object & NGC\,6905 & 20:22:22.6 & 20:06:15.5     & 2$\times$1200  &1.15\\
       &stds   & Wolf\,1346& 20:34:21.2 & 25:03:34.4     & 6$\times$60    &1.53\\
       &arc    & HeNe      & --         & --             & 2$\times$2     &1.35\\
       &flat   & Halogen   & --         & --             &10$\times$33.3  &1.00\\
       &bias   & --        & --         & --             &24$\times$0     &--  \\
\hline
\end{tabular}
\label{tab:obs} 
\end{center}
\end{table}

\subsection{IR data}

In order to produce a consistent model of the nebular and dust properties of NGC\,6905,
we also need the infrared (IR) photometry of this WRPN.
The necessary IR data were retrieved from the NASA/IPAC IR archive\footnote{\url{https://irsa.ipac.caltech.edu/frontpage/}}.
Specifically, we retrieved archival IR images obtained from {\itshape Spitzer}, {\itshape WISE}, {\itshape IRAS} and {\itshape Akari}. With these, we can create a spectral energy distribution (SED) that covers the 5.8--160~$\mu$m spectral range. Details of the observations used here are listed in Table~\ref{tab:table1}.

\begin{table}
\begin{center}
\caption{IR observations of NGC 6905.}
\setlength{\tabcolsep}{0.8\tabcolsep}    
\begin{tabular}{lcccc}
\hline
Telescope     & Instrument & Band      & Obs.\,date    & Obs.\,ID.      \\
              &            & ($\mu$m)  &  (yyyy-mm-dd) &                \\
\hline
{\itshape Spitzer} & IRAC       & 5.8, 8.0  & 2004-10-08    & 4421376        \\
{\itshape WISE}    &            & 12, 22    & 2010-05-01    & 3056p196\_ac51 \\
{\itshape IRAS}    & ISSA-II    & 60, 100   & 1992-04-22    &                \\
{\itshape Akari}   & FIS        & 140, 160  & 2000-01-01    &                \\
\hline
\end{tabular}
\label{tab:table1}
\end{center}
\end{table}

\begin{table}
\begin{center}
\caption{IR and radio photometric fluxes of NGC 6905.}
\setlength{\tabcolsep}{\tabcolsep}    
\begin{tabular}{lccc}
\hline
Wavelength & Band     & Flux  & Error\\
range      & ($\mu$m) & (mJy) & (mJy)\\
\hline     
Infrared   & 5.8      & 45    & 1 \\
           & 8.0      & 165   & 1 \\
           & 12       & 652   & 26 \\
           & 22       & 3471  & 19 \\
           & 60       & 7620  & 71 \\
           & 100      & 7600  &2800 \\
           & 140      & 2730  &280 \\
           & 160      & 1920  &560 \\
\hline
           & (cm)       &       &   \\
Radio      & 6    & 63    & 10 \\
           & 20   & 67    & 3 \\
\hline
\end{tabular} 
\label{tab:ir_photometry}
\end{center}
\end{table}

IR photometric measurements were extracted from each image. We defined an extraction region that
encompasses the nebular emission for each individual IR image.
The corresponding photometric measurements were calculated adding up all the flux from the pixels within this region. Several background regions were selected from the vicinity of NGC\,6905 with no contribution from the nebula, additionally we excised the contribution of background stars like the south-east star at the edge of the nebula.
In Table~\ref{tab:ir_photometry} we list the photometric values and their errors.
The errors account for the calibration uncertainty from each instrument as well as the standard deviation from the selected background regions. Details of this process can be found in  \citet{Jimenez2020,Jimenez2021} and \citet{Rubio2020} and will not be discussed here. We do not include the shorter wavelengths from {\itshape Spitzer} because of the presence of numerous background stars that hinders a proper photometry extraction.

Finally, photometric measurements obtained at radio frequencies were also included.
Radio measurements at 1.4 and 4.85~GHz, that correspond to 20 and 6 cm,
respectively, were also taken from the NASA/IPAC infrared archive, corresponding to the National Radio Astronomy Observatory (NRAO) Very Large Array (VLA) survey and the NRAO Green Bank telescope (GBT), respectively \citep[see also][]{Hajduk2018}.
The radio measurements are also listed in Table~\ref{tab:ir_photometry}.

\section{ANALYSIS} 

The NOT ALFOSC spectrum of HD\,193949 presented in Figure~\ref{fig:spec_g7} corroborates its [WR] nature with the clear presence of the broad features at $\sim$4686~\AA\, and $\sim$5806~\AA, the so-called blue and red bumps (hereinafter BB and RB), together with other WR features. The CSPN of NGC\,6905 also exhibits the broad \ovi\ feature at
$\sim$3820~\AA\, (hereinafter the violet bump or simply VB), which has been used to classify this object as part of the oxygen sequence \citep{Acker2003}.
Additionally, twenty one WR emission lines can be spotted in the 
ALFOSC spectrum at different wavelengths between 3680~\AA{} and 7060~\AA{}, plus several high-excitation Ne lines, which have never been discussed before.
Several contributing narrow emission lines of nebular origin (non-stellar) are also labeled on the spectrum.
All the ions responsible for the broad emission lines displayed in the spectrum of HD\,193949 are listed in 
Table~\ref{tab:elines}.
Most of the observed wavelengths of the ions analysed here were consulted in the National Institute for Standards and Technology (NIST)
Atomic Spectra Database\footnote{\url{https://physics.nist.gov/PhysRefData/ASD/lines_form.html}}.

Previous authors have developed a methodology to sub-classify [WR]-type CSPNe using the
line fluxes and/or the equivalent widths (EW) of the broad spectral emission features 
\citep[e.g.,][]{Acker2003,Crowther1998}.
For this, we need to calculate the line fluxes for all the WR features in the spectrum; we note that some broad features are usually the result of blends of stellar broad features with some contribution from nebular
lines. Although minor, these need to be taken into account. To help us disentangle the presence from the contributing nebular lines to the WR bumps, we applied the multi-Gaussian fitting
technique discussed in length in Paper~I. This method allows us to assess the contribution from
nebular lines (e.g.: \ariv\,$\lambda4711$ and \heiiwr\ in the BB).
We give a brief description in the following.

\begin{figure*}
\begin{center}
\includegraphics[angle=0,width=0.33\linewidth]{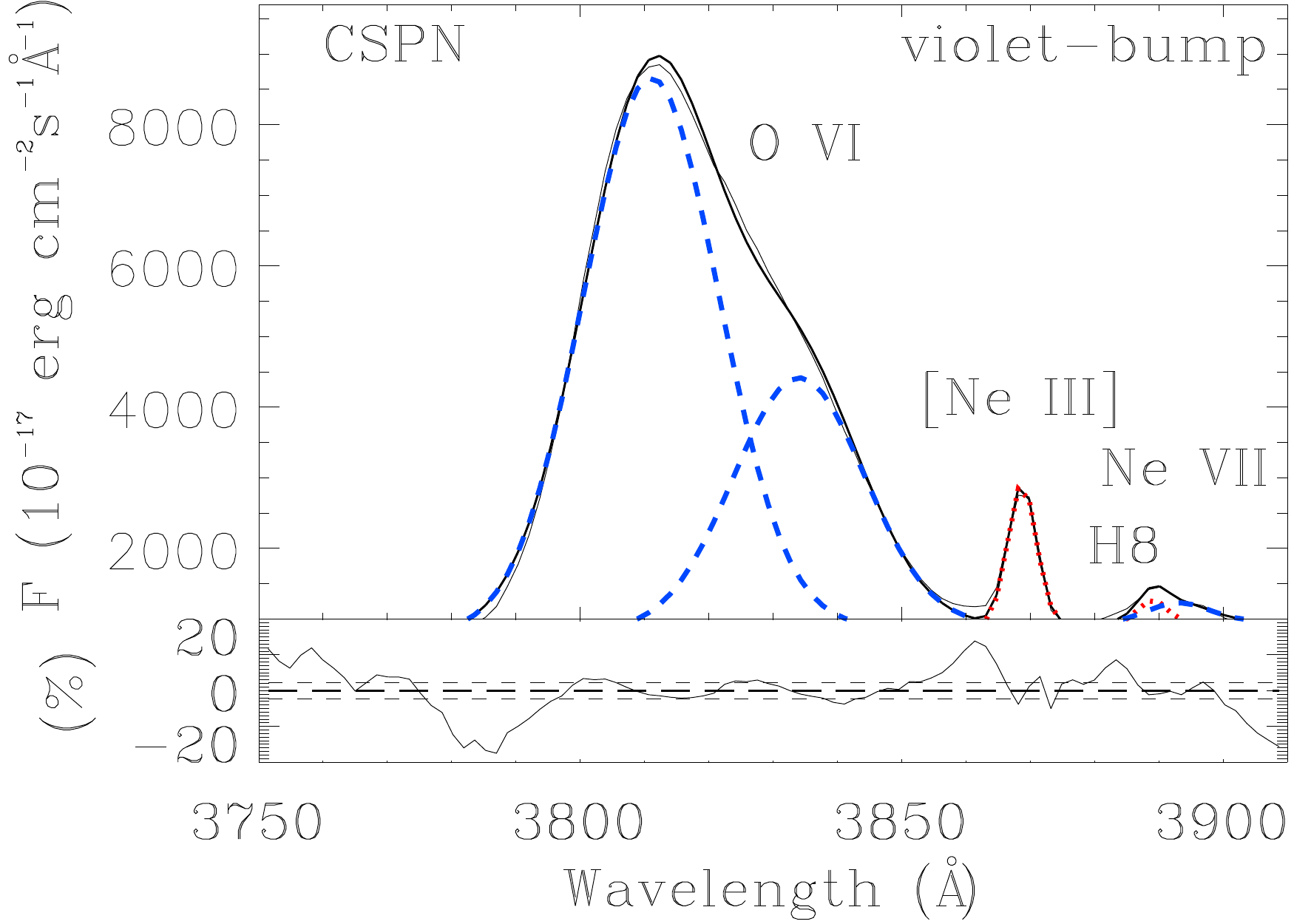}~
\includegraphics[angle=0,width=0.33\linewidth]{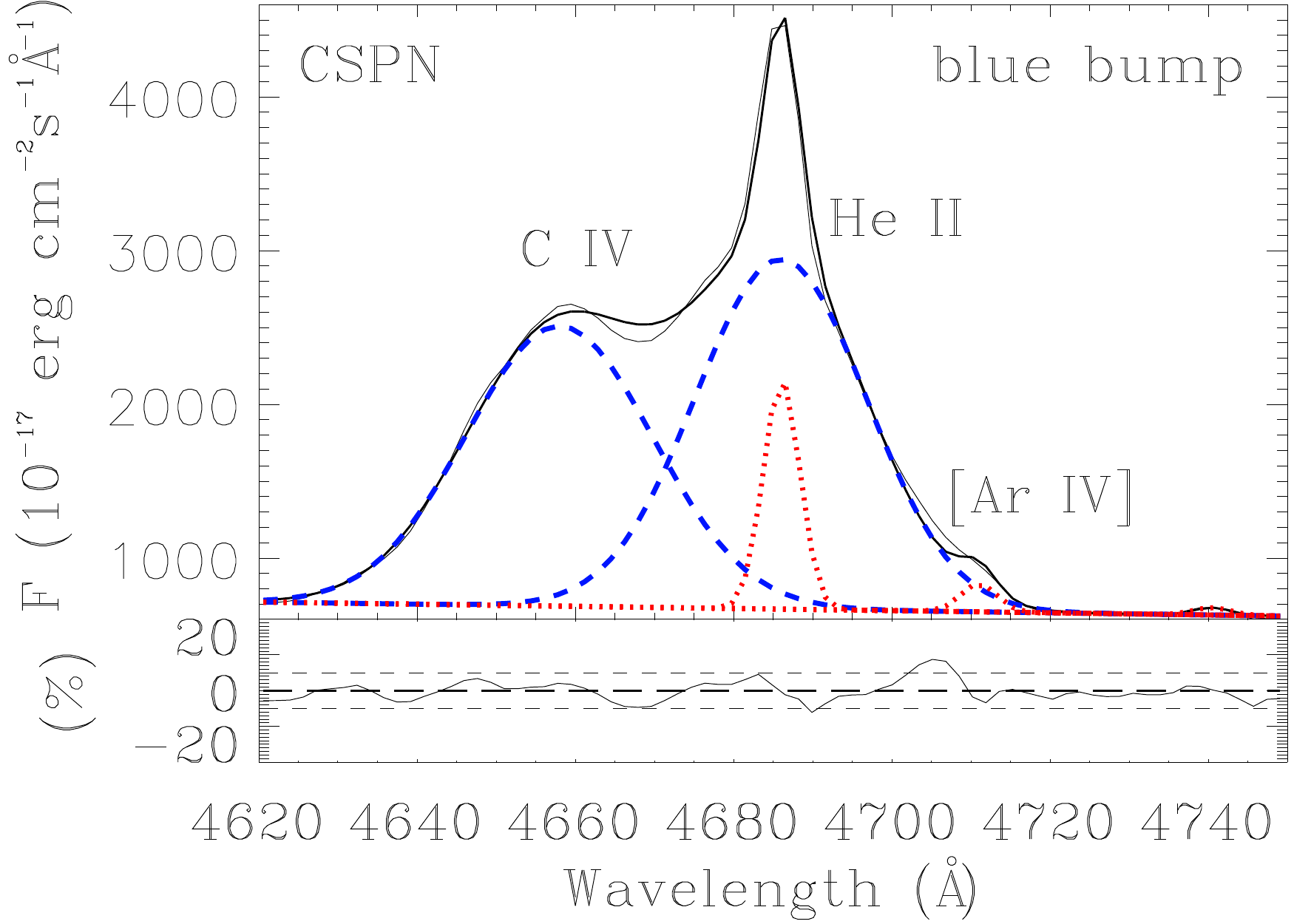}~
\includegraphics[angle=0,width=0.33\linewidth]{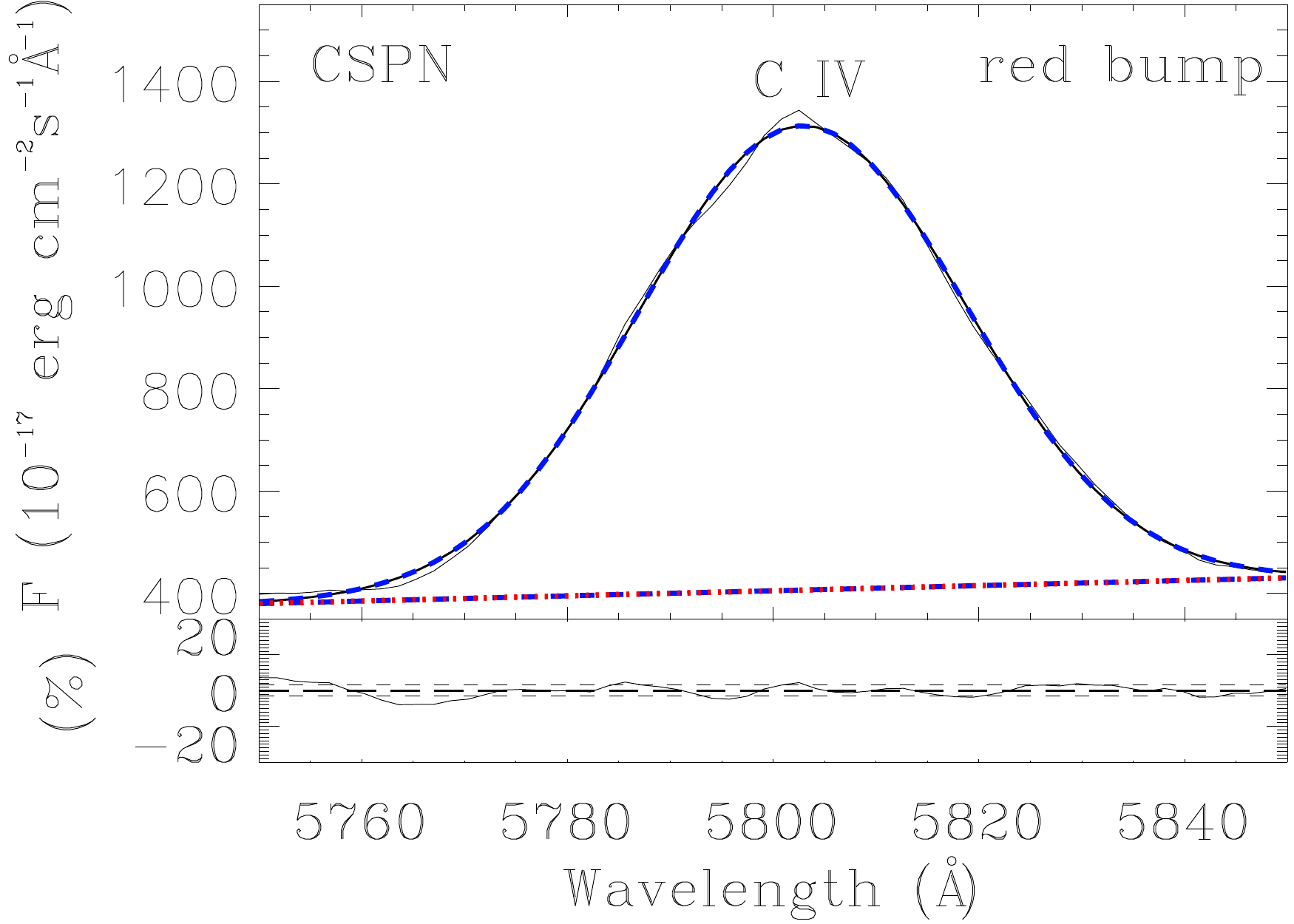}\\
\includegraphics[angle=0,width=0.33\linewidth]{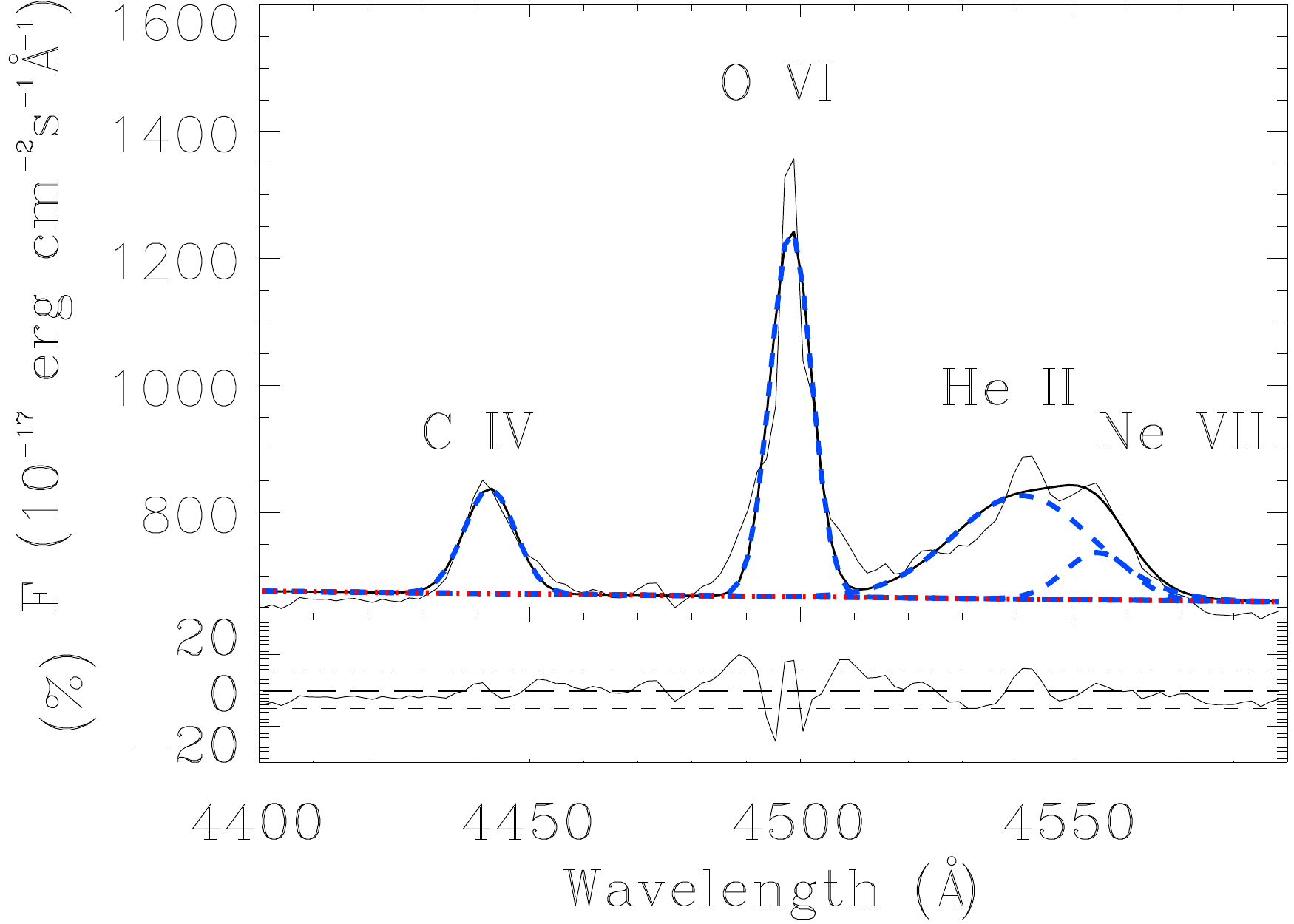}~
\includegraphics[angle=0,width=0.33\linewidth]{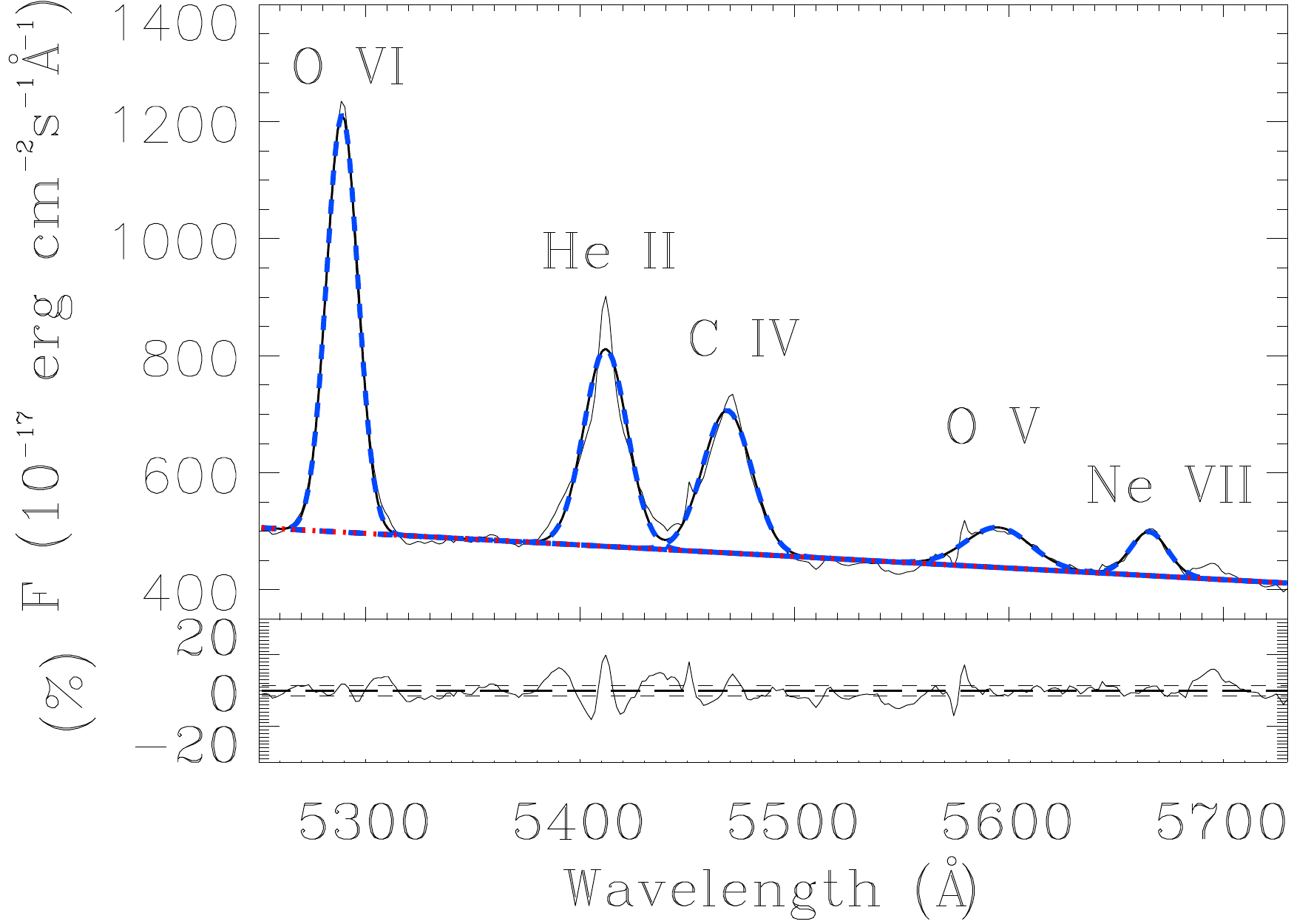}~
\includegraphics[angle=0,width=0.33\linewidth]{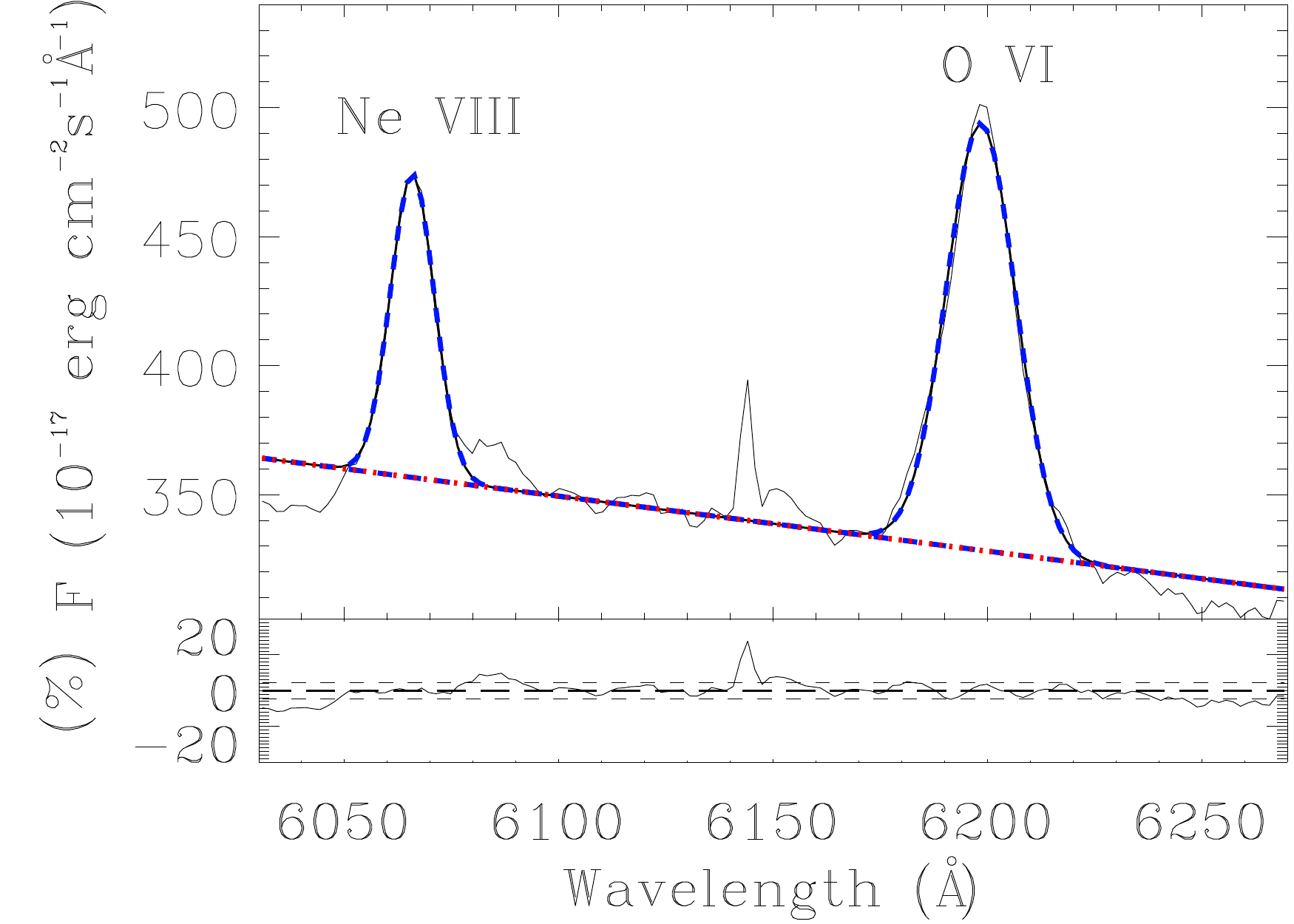}\\
\includegraphics[angle=0,width=0.33\linewidth]{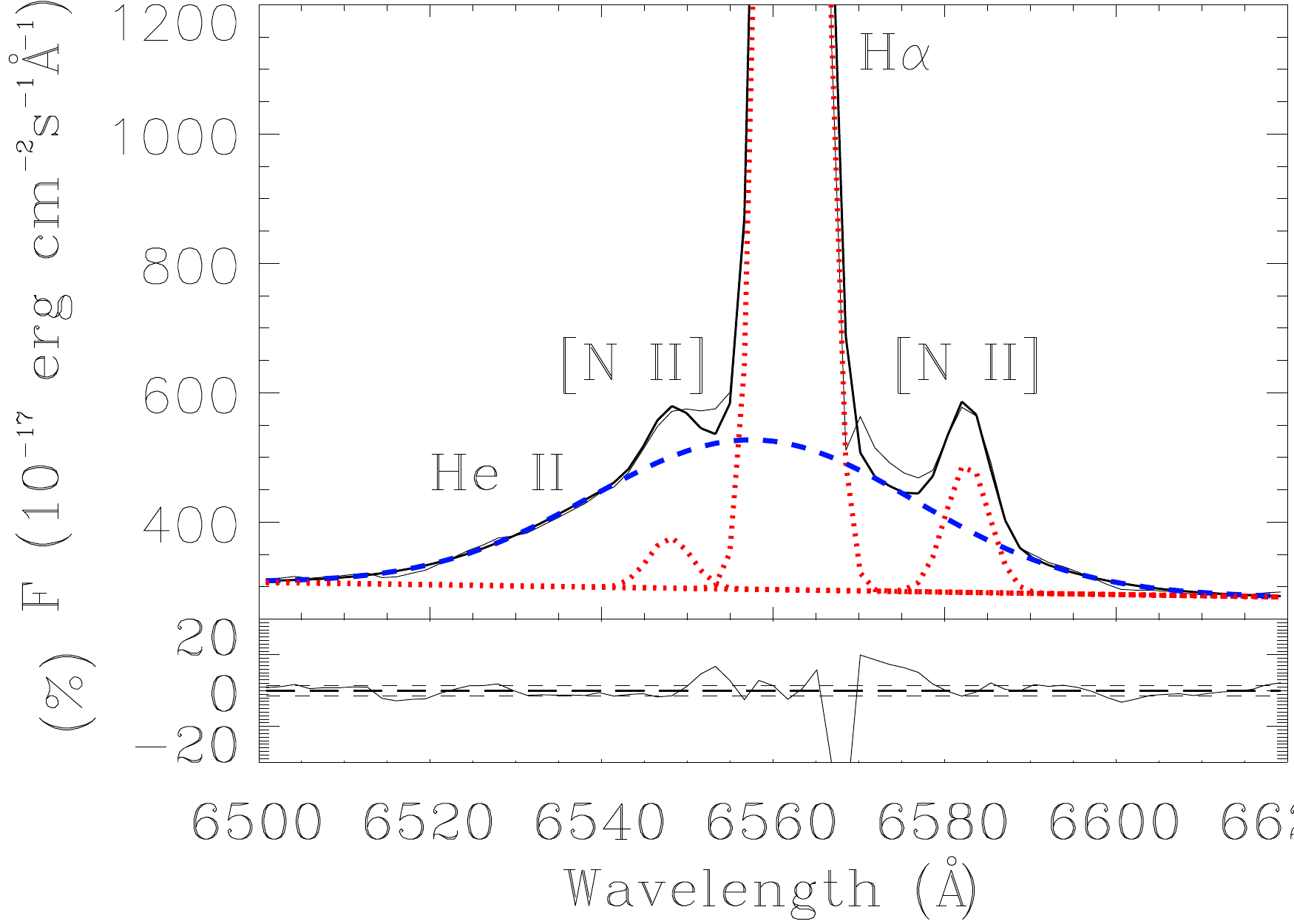}~
\includegraphics[angle=0,width=0.33\linewidth]{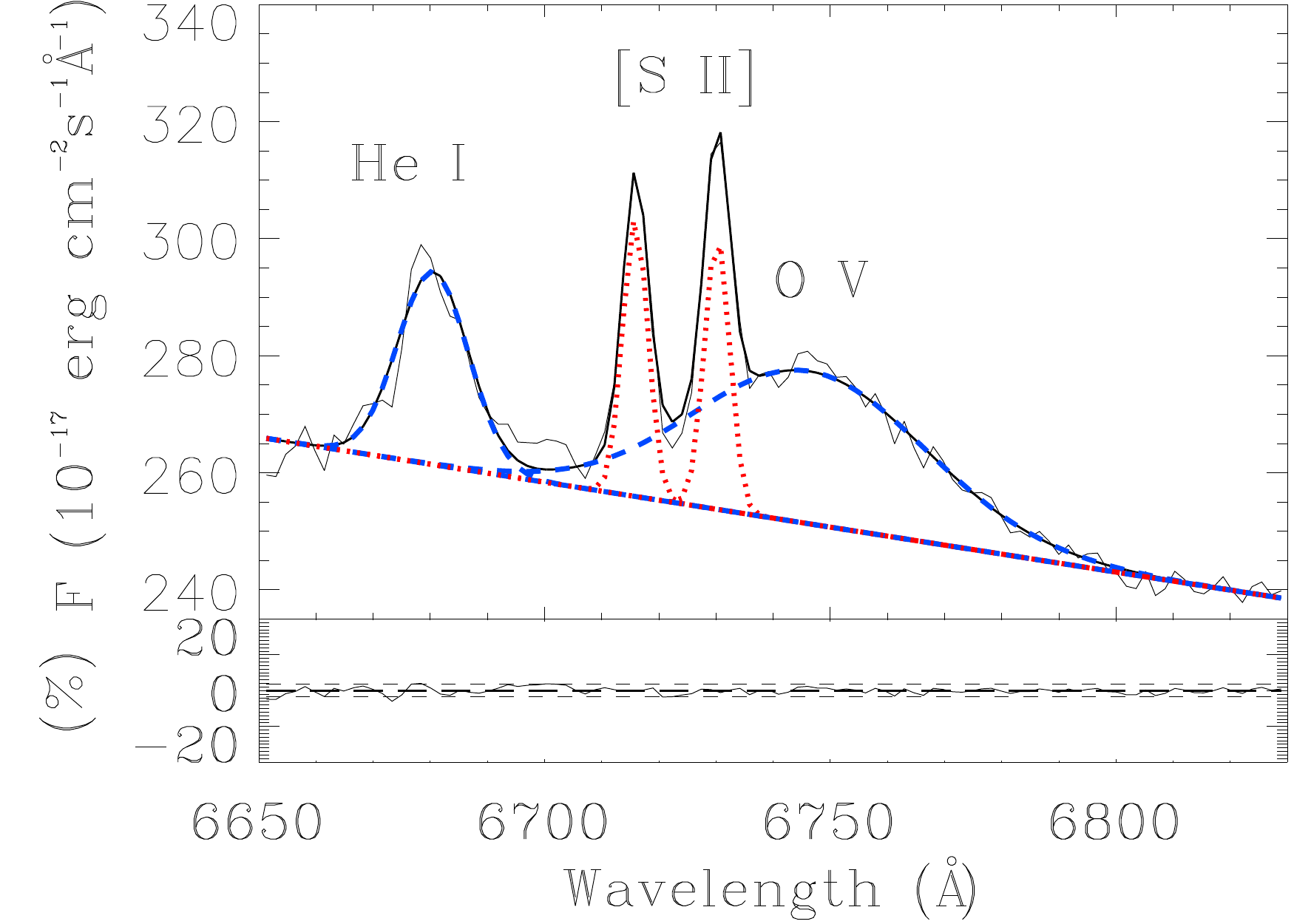}~
\includegraphics[angle=0,width=0.33\linewidth]{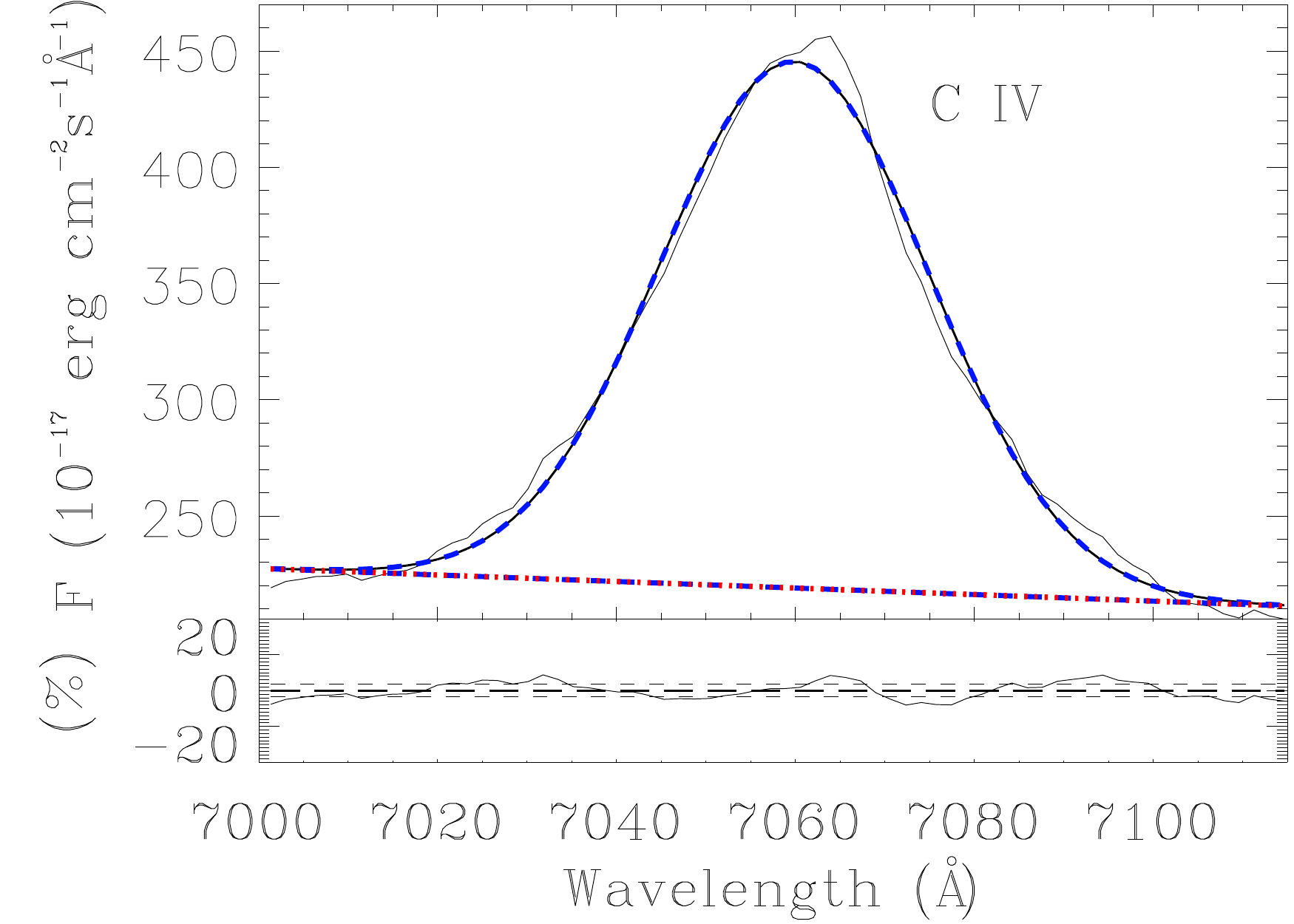}
\caption{Multicomponent Gaussian fits to the bumps and other WR features of the CSPN on NGC\,6905:
VB ({\itshape top left}); BB ({\itshape top center}) and RB ({\itshape top right}).
The rest of the panels show other WR features and high ionization Ne emission lines (see Fig.~\ref{fig:spec_g7}). 
The blue dashed lines represents the fits to the broad WR features and the red dotted-lines the narrow nebular emission. The sum is shown in black. The fitted continuum is shown by the dashed straight line. The residuals are shown at the bottom of each panel (in per cent).
The results are listed in Table~\ref{tab:elines}.}
\label{fig:WR_features}
\end{center}
\end{figure*}

\subsection{Line fluxes of the CSPN HD\,193949}

\begin{table*}
\begin{center}
\caption{Parameters of the WR features in the optical spectrum of the CSPN of NGC\,6905.}
\setlength{\tabcolsep}{0.75\tabcolsep}
\begin{tabular}{cccccccccccc}
\hline
ID & Nature & Ion & $\lambda_0$ & $\chi$ & $\lambda_\mathrm{c}$ & $F^{\dagger}$     & FWHM & $EW$  & SNR  & $F_\mathrm{\lambda}/F_\mathrm{RB}$ & $EW_\mathrm{\lambda}/EW_\mathrm{RB}$\\
   &     &        & (\AA)& (eV)&(\AA)&($10^{-14}$~\ergcms)&(\AA) & (\AA)    &       & ($F_\mathrm{RB}=100$)& ($F_\mathrm{RB}=100$) \\
(1)& (2) & (3)    & (4)  & (5)   & (6)    & (7)            & (8)  & (9)          & (10)  & (11)            & (12)             \\
\hline
VB &     &        &      &       &        &                &      &              &       &                 &                 \\
1  & WR  & \ovi\  & 3811 & 138.1 & 3811.0 &206.99$\pm$0.46 & 24.8 &245.9$\pm$1.3 &  43.6 & 572.74$\pm$2.29 & 279.37$\pm$0.15 \\
2  & WR  & \ovi\  & 3834 & 138.1 & 3834.0 & 90.25$\pm$0.36 & 23.9 &102.7$\pm$0.6 &  43.6 & 249.72$\pm$1.30 & 116.69$\pm$0.19 \\
BB &     &        &      &       &        &                &      &              &       &                 &                 \\
3  & WR  & \civ\  & 4658 &  64.5 & 4658.0 & 53.95$\pm$0.59 & 27.8 & 77.1$\pm$1.1 &  20.2 & 149.28$\pm$1.71 &  87.55$\pm$0.86 \\
4  & WR  & \heii\ & 4686 &  24.6 & 4686.0 & 61.52$\pm$0.58 & 25.4 & 90.4$\pm$1.2 &  20.2 & 170.23$\pm$1.70 & 102.71$\pm$0.94 \\
5  & Neb & \heii\ & 4686 &  24.6 & 4686.1 &  8.62$\pm$0.24 &  5.4 & 12.7$\pm$0.4 &  20.2 &  23.85$\pm$0.67 &  14.39$\pm$0.45 \\
6  & Neb & \ariv\ & 4711 &       & 4711.0 &  1.05$\pm$0.19 &  5.4 &  1.6$\pm$0.3 &  20.2 &   2.91$\pm$0.53 &   1.80$\pm$0.32 \\
RB &     &        &      &       &        &                &      &              &       &                 &                 \\
7  & WR  & \civ\  & 5806 &  64.5 & 5802.8 & 36.14$\pm$0.12 & 37.4 & 88.0$\pm$0.4 &  65.5 & 100.00$\pm$0.47 & 100.00$\pm$0.03 \\
other lines & & &  &       &        &                &     &              &       &                 &                 \\
8  & WR  & \civ\  & 3689 &  64.5 & 3689.0 &  5.06$\pm$0.20 & 14.4 &  5.7$\pm$0.2 &  43.6 &  14.00$\pm$0.56 &   6.50$\pm$0.22 \\
9  & WR  & \civ\  & 3933 &  64.5 & 3932.4 &  3.19$\pm$0.19 & 18.1 &  4.2$\pm$0.2 &  43.6 &   8.83$\pm$0.53 &   4.73$\pm$0.25 \\
10 & WR  & \heii\ & 4200 &  24.6 & 4200.6 &  1.29$\pm$0.17 & 17.4 &  1.8$\pm$0.2 &  43.6 &   3.57$\pm$0.47 &   2.04$\pm$0.25 \\
11 & WR  & \civ\  & 4227 &  64.5 & 4226.7 &  2.96$\pm$0.18 & 18.9 &  4.2$\pm$0.2 &  43.6 &   8.19$\pm$0.50 &   4.72$\pm$0.26 \\
12 & WR  & \civ\  & 4440 &  64.5 & 4442.8 &  1.95$\pm$0.28 & 11.0 &  2.9$\pm$0.4 &  20.2 &   5.40$\pm$0.77 &   3.30$\pm$0.45 \\
13 & WR  & \ovi\  & 4500 & 138.1 & 4498.1 &  5.48$\pm$0.27 &  8.9 &  8.2$\pm$0.4 &  20.2 &  15.16$\pm$0.75 &   9.34$\pm$0.43 \\
14 & WR  & \heii\ & 4541 &  24.6 & 4541.0 &  5.19$\pm$0.45 & 29.9 &  7.8$\pm$0.7 &  20.2 &  14.36$\pm$1.25 &   8.89$\pm$0.73 \\
15 & WR  & \civ\  & 4789 &  64.5 & 4789.2 &  1.99$\pm$0.33 & 21.4 &  3.4$\pm$0.6 &  20.2 &   5.51$\pm$0.91 &   3.84$\pm$0.62 \\
16 & WR  & \ov\   & 4933 & 113.9 & 4933.1 &  4.87$\pm$0.38 & 30.3 &  8.7$\pm$0.7 &  20.2 &  13.48$\pm$1.05 &   9.92$\pm$0.74 \\
17 & WR  & \ov\   & 5100 & 113.9 & 5100.0 &  3.60$\pm$0.41 & 41.5 &  6.9$\pm$0.8 &  20.2 &   9.96$\pm$1.14 &   7.84$\pm$0.86 \\
18 & WR  & \ovi\  & 5166 & 138.1 & 5166.5 &  0.52$\pm$0.06 & 11.4 &  1.0$\pm$0.1 &  20.2 &   1.44$\pm$0.17 &   1.15$\pm$0.13 \\
19 & WR  & \ovi\  & 5290 & 138.1 & 5288.9 & 13.13$\pm$0.09 & 17.3 & 26.4$\pm$0.2 &  66.8 &  36.33$\pm$0.28 &  29.94$\pm$0.09 \\
20 & WR  & \heii\ & 5412 &  24.6 & 5411.8 &  8.49$\pm$0.09 & 23.6 & 17.9$\pm$0.2 &  66.8 &  23.49$\pm$0.26 &  20.34$\pm$0.13 \\
21 & WR  & \civ\  & 5470 &  64.5 & 5468.6 &  6.89$\pm$0.10 & 26.6 &114.9$\pm$0.2 &  66.8 &  19.07$\pm$0.26 &  16.91$\pm$0.15 \\
22 & WR  & \ov\   & 5600 & 113.9 & 5595.0 &  2.59$\pm$0.07 & 35.5 &  5.9$\pm$0.2 &  65.5 &   7.17$\pm$0.28 &   6.72$\pm$0.22 \\
23 & WR  & \ovi\  & 6200 & 138.1 & 6198.5 &  3.35$\pm$0.08 & 19.0 & 10.2$\pm$0.3 &  43.3 &   9.27$\pm$0.22 &  11.61$\pm$0.24 \\
24 & WR  & \heii\ & 6406 &  24.6 & 6406.0 &  0.63$\pm$0.03 & 11.8 &  2.2$\pm$0.2 &  43.3 &   1.74$\pm$0.14 &   2.51$\pm$0.20 \\
25 & WR  & \heii\ & 6560 &  24.6 & 6557.7 & 10.95$\pm$0.07 & 44.5 & 36.7$\pm$0.4 &  43.3 &  30.30$\pm$0.35 &  41.69$\pm$0.29 \\
26 & WR  & \ov\   & 6740 & 113.9 & 6746.3 &  1.33$\pm$0.10 & 47.9 &  5.3$\pm$0.4 &  43.3 &   3.68$\pm$0.26 &   6.01$\pm$0.40 \\
27 & WR  & \civ\  & 7060 &  64.5 & 7059.9 &  8.54$\pm$0.04 & 35.4 & 39.0$\pm$0.3 &  57.9 &  23.63$\pm$0.18 &  44.31$\pm$0.15 \\
special lines$^{\dagger}$$^{\dagger}$  & & &  &       &        &                &                &                   &       &                 &                 \\
28 & stellar emission &\nevii\  & 4555 & 207.3 & 4552.5 & 1.14$\pm$0.30 & 14.4 &  1.7$\pm$0.5 &  20.2 &   3.15$\pm$0.83 &   1.96$\pm$0.51 \\
29 & stellar emission &\nevii\  & 5666 & 207.3 & 5665.4 & 1.63$\pm$0.07 & 20.4 &  3.8$\pm$0.2 &  65.5 &   4.51$\pm$0.19 &   4.36$\pm$0.17 \\
30 & stellar emission &\neviii\ & 6068 & 239.1 & 6065.9 & 1.51$\pm$0.07 & 12.1 &  4.2$\pm$0.2 &  43.3 &   4.18$\pm$0.19 &   4.82$\pm$0.21 \\

\hline
\end{tabular}\\
{\itshape Notes}.
(1) Identification number;
(2) nature of the contributing emission line: WR (broad/stellar) or nebular (narrow);
(3) ion;
(4) rest wavelength in \AA;
(5) Ionisation potential [eV];
(6) observed centre of the line;
(7) flux [$10^{-14}$~\ergcms];
(8) full width at half maximum (FWHM) [\AA];
(9) equivalent width (EW) [\AA];
(10) continuum signal-to-noise ratio (SNR) closest to the feature;
(11) line fluxes normalized to that of the RB. The ratio has been computed adopting $F$(RB)=100.
(12) EW normalized to that of the RB, adopting $EW$(RB)=100.
Special lines$^{\dagger}$$^{\dagger}$ correspond to high ionization Ne lines
from stellar emission origin \citep{Werner2007}.
$^{\dagger}$ Only the statistical errors, determined with equations 1 and 2, are shown in this table.
\label{tab:elines}
\end{center}
\end{table*}

\begin{table*}
\begin{center}
\caption{Diagnostic criteria to classify HD\,193949 as a [WR]-type CSPN.}
\begin{tabular}{cccccc}
\hline

Criteria                                    & [WO1]         & [WO2]      & [WO3]     & WRPN & Classification \\
(1)                                          & (2)           & (3)        & (4)       & (5)  & (6)            \\
\hline
\citet[][]{Acker2003}\\
$F_\text{\ovi\ 3822}/F_\text{RB}$            & $>1400$       &1000$\pm$200&250$\pm$40 & 822.5$\pm$2.6  & [WO2]    \\
$F_\text{\civ\ 4650}/F_\text{RB}$            &300$\pm$100    &270$\pm$60  & 23$\pm$2  & 149.3$\pm$1.7  & [WO2]    \\
$F_\text{\civ\ 4686}/F_\text{RB}$            &500$\pm$200    &300$\pm$30  &130$\pm$30 &170.2$\pm$1.7   & [WO2--3] \\
$F_\text{\ovi\ 5290}/F_\text{RB}$$^{\dagger}$          &   $>80$       & 48$\pm$2   & 20$\pm$5  & 36.3$\pm$0.3   & [WO2--3] \\
$F_\text{\civ\ 5412}/F_\text{RB}$            & 45$\pm$15     & 20$\pm$4   & 15$\pm$4  & 23.5$\pm$0.3   & [WO2]    \\
$F_\text{\civ\ 5470}/F_\text{RB}$            & 35$\pm$5      & 23$\pm$2   & 14$\pm$2  & 19.1$\pm$0.3   & [WO2]    \\
$F_\text{\nevii\ 5666}/F_\text{RB}$$^{\dagger}$          &   $>25$       &  10:       &  8$\pm$6  &  4.5$\pm$0.2   & [WO2--3] \\
$F_\text{\neviii\ 6068}/F_\text{RB}$$^{\dagger}$         &   20$\pm$8    &  6$\pm$1   &  2$\pm$1  & 4.18$\pm$0.2   & [WO2--3] \\
$F_\text{\ov\ 6740}/F_\text{RB}$             & presence      & --         & --        &  3.7$\pm$0.3   & [WO1]    \\
$F_\text{\civ\ 7060}/F_\text{RB}$            & 35$\pm$20     & 18$\pm4$   & 15$\pm$4  & 23.6$\pm$0.2   & [WO1--2] \\
FWHM of \civwrrb [\AA]                       &   33$\pm$5    & 32$\pm$3   & 37$\pm$6  & 37.4           & [WO1--3] \\
\hline
\citet[][]{Crowther1998}\\
log$(EW_\text{\ovi3820}/EW_\text{\ov5590})$  & $\geq1.1$     & 0.6--1.1   & 0.25--0.6 & 1.58           & [WO1]    \\
log$(EW_\text{\ovi3820}/EW_\text{\civ5806})$ & $\geq0.2$     & $\geq0.2$  & $-1$--0.2 & 0.46           & [WO1--2] \\
log$(EW_\text{\nevii5666}/EW_\text{\ov5590})$$^{\dagger}$  & $\geq0.0$     & $\leq0.0$  & $<<0.0$   & $-0.19$        & [WO2]    \\
FWHM of \civwrrb\ [\AA]                      & 40$\pm$10     &160$\pm$20  & 90$\pm$30 & 37.4           & [WO1]    \\

\hline
\end{tabular}\\
\label{tab:class}
\end{center}
$^{\dagger}$Note that we here identify the $\lambda5290$ emission line as \ovi\ and not as \cvi; the $\lambda5666$ as \nevii\ and not as \ovii; and the $\lambda6068$ as \neviii\ and not as \oviii, as previously defined \citep[see][]{Crowther1998,Acker2003}. However, according to spectral observations  these updated criteria remains as valid as before \citep[see][]{Werner2007}, and given its importance for classification, the re-identification is worth clarifying for future use. See text for details.
\end{table*}

The WR features in the spectrum of the CSPN HD\,193949 (Figure~\ref{fig:spec_g7}) were decomposed into
their individual emission lines by following the multi-Gaussian
approach \citep[discussed in length in][]{Gomez2020b}.
In brief, the method consists of fitting the individual WR features with
multi-Gaussian components using a tailor-made code that
uses the {\scshape idl} routine {\scshape lmfit} which performs a non-linear least squares fit
of a function with an arbitrary number of parameters by using the Levenberg-Marquardt algorithm  \citep{Press1992}.
The resultant fits to each broad spectral feature give us the line fluxes ($F$), 
central wavelength ($\lambda_\mathrm{c}$), full width at half-maximum (FWHM) and EW from
the contributing lines from stellar and nebular origin.
Table~\ref{tab:elines} resumes our findings.

Examples of multi-Gaussian fitting to several WR features in HD\,193949 are illustrated in Figure~\ref{fig:WR_features}. In order of increasing ionization potential, the WR features detected are:
\heii~$\lambda\lambda$4200, 4541, 4686, 5412, 6406, 6560;
\civ~$\lambda\lambda$3689, 3933, 4227, 4440, 4658, 4789, 5470, 5806, 7060;
\ov~$\lambda\lambda$4933, 5100, 5600, 6740 and
\ovi~$\lambda\lambda$3811, 3834, 4500, 4604, 5166, 5290, 6200.
We highlight the presence of three Ne broad emission lines: \nevii~$\lambda\lambda4555,5666$ and
\neviii~$\lambda6068$, of stellar origin according to \citet{Werner2007}. These lines are discussed in the following sections.

We found that the VB is composed of two broad emission lines 
of \ovi~$\lambda\lambda3811,3834$, with negligible contribution of \neiii\ and \nevii\
at its red wing (see Fig.~\ref{fig:WR_features} top left panel). 
The BB, composed by the broad \heii~$\lambda4686$ and \civ~$\lambda4658$ features, has some contribution from nebular lines of \heii\ and \ariv\ (Fig.~\ref{fig:WR_features} top middle panel). 
The RB is fitted by a single broad line at $\sim\lambda5806$~\AA, from the blended components of \civ~$\lambda\lambda5801,5812$ (Fig.~\ref{fig:WR_features} top right panel).
Notice that they all correspond to broad emission lines (FWHM$>$5.4 \AA; see Table~\ref{tab:elines}, column~8), according to their WR nature.

The statistical errors of the emission lines were determined as the 1-$\sigma$ deviation, $\sigma_l$, on each measured flux of the emission line using the expressions from \cite{Tresse1999}:
\begin{equation}
\sigma_l = \sigma_c D \sqrt{2 N_{\rm pix} + \frac{EW}{D}},
\end{equation}
\noindent and for the EW:
\begin{equation}
\sigma_{EW} = \frac{EW}{F} \sigma_c D \sqrt{\frac{EW}{D} + 2 N_{\rm pix} + \left(\frac{EW}{D}\right)^{2}},
\end{equation}
\noindent where $D$=1.7~\AA~pixel$^{-1}$ is the spectral dispersion (for NOT ALFOSC), $\sigma_c$
is the mean
standard deviation per pixel of the continuum and $N_{\rm pix}$ is the number of
pixels covered by the feature.

In Table~\ref{tab:elines} we also list what we call "special lines",
corresponding to high ionization potential Ne lines (see column 5)
from presumably stellar emission origin.
Before \citet{Werner2007}, the \neviii\ emission line at $\sim$6068~\AA,
with a ionization potential of 239.1~eV, was previously identified as a non-stellar
ultra-high ionization O feature, \oviii, with a much higher ionization potential of 871.4~eV. This ion supposedly formed in shock wind regions.

The same applied to other spectral features like \oviii~$\lambda4340$ now identified as \neviii;
\cvi~$\lambda4500$ now identified as \ovi; \ovii~$\lambda4555$ now identified as \nevii;
\cv~$\lambda4945$ now identified as \nv; \cvi~$\lambda5290$ now identified as \ovi\ and
\ovii~$\lambda5665$ now identified as \nevii.
We note that although the presence of X-ray-emitting gas is invoked in low $T_\mathrm{eff}$ CSPN to explain the ultrahigh-ionization O features \citep[see, e.g. the case of IC\,4593;][]{Herald2011,Toala2020}, this might not be the case for hot WRPN.
Furthermore, \citet{Werner2007} showed that the mere presence of \nevii\ and \neviii\
puts a strict lower limit of $T_\mathrm{eff} \approx 150$~kK.

In Table~\ref{tab:elines} we present the flux, FWHM, {\it EW}, as well as the parameters $F_\mathrm{\lambda}/F_\mathrm{RB}$ adopting $F_\mathrm{RB}=100$ and $EW_\mathrm{\lambda}/EW_\mathrm{RB}$ adopting $EW_\mathrm{RB}=100$, that are used as a classification criteria for CSPNe, according with \citet[][]{Acker2003} and \citet{Crowther1998}. The mere presence of the strong \ovi\ broad emission line $\sim\lambda3820$~\AA, the VB in the CSPN spectrum of NGC\,6905 (see Fig.~\ref{fig:spec_g7}), suggests a [WO]-type classification.
Moreover, the presence of high-ionized Ne lines constrains it to a very early-type \citep[e.g][]{Werner2007}.
A classification of CSPNe later than [WO2]-subtypes can be safely discarded since they do not display neither \nevii\ nor \neviii\ in their spectra.
Besides, later [WO]-subtypes, like [WO4], present weaker \ovi\ and \civ\ intensities, in general.
Also, any [WC] fingerprint (early or later) is also discarded:
\ciii\ lines, expected in [WCL]-subtypes, are absent in our spectrum, and in the classification schemes by \citet[][]{Acker2003} and \citet[][]{Crowther1998}, [WCE]-subtypes do not display a strong VB, by definition. It is worth to explicitly mention that we also did not detect any broad N lines, which also rules out any [WN]-subtype or transitional stages [WN/C].

In order to quantitatively classify the CSPN of NGC\,6905 as [WO]-type, Table~\ref{tab:class} summarizes the values of the main parameters among the [WO]-subtypes in which our [WR] star is expected to be ([WO1], [WO2] and [WO3]), according to the lines present in our spectrum. The spectrum of HD\,193949 suggest a [WO2]-subtype based on the \citet[][]{Acker2003} criteria, whilst the \citet{Crowther1998} scheme suggest a [WO1]-subtype. In conclusion, we classified the CSPN of NGC\,6905 as a [WO]-type, specifically an early one, between [WO1] and [WO2]-subtype.

\subsection{NLTE analysis of HD\,193949}

We have analysed UV and optical spectra by means of the most updated version (2021 January 24) of the non-local thermodynamic equilibrium (NLTE) stellar atmosphere code Potsdam Wolf-Rayet \citep[PoWR;][]{Grafener2002,Hamann2004}\footnote{\url{http://www.astro.physik.uni-potsdam.de/~wrh/PoWR}} in order to infer the stellar parameters of HD\,193949.
Details of the computing scheme can be found in \citet{Todt2015}.

Available UV \emph{Far Ultraviolet Spectroscopic Explorer} (\emph{FUSE}) and {\it Hubble Space Telescope (HST)} Space Telescope Imaging Spectrograph (STIS) observations of HD\,193949 were retrieved from the Mikulski Archive for Space Telescopes (MAST)\footnote{\url{https://archive.stsci.edu/hst/}}.
The UV observations were analysed in conjunction with the optical NOT ALFOSC spectra with the PoWR stellar atmosphere code. The {\it FUSE} observation corresponds to Obs.\ ID.\ a1490202000 (PI: F.\ Bruhweiler) and were obtained on 2000 August 11 with the LWRS aperture for a total exposure time of 9487\,s. The {\it HST} STIS data were taken on 1999 June 29 and correspond to Obs.\ ID.\ o52r01010 (NUV-MAMA) and o52r01020 (FUV-MAMA), each with an exposure time of 802\,s (PI: A.\ Boggess) and taken with the time-tag mode with the $52 \times 0.5\arcsec$ aperture. The FUV-MAMA observation was performed with the G140L grating at a central wavelength of 1425\,\AA{}, and the NUV-MAMA with the G230L grating at a central wavelength of 2376\,\AA{}. We notice that there might be a problem with the background subtraction for the FUV-MAMA observation, as the flux at the centre of the interstellar Lyman-$\alpha$ absorption line as well as the absorption trough of the \civ\ resonance doublet does not reach zero.  

As the typical emission-line spectra of WR-type stars are predominantly formed by recombination processes in their dense
stellar winds, the continuum-normalized spectrum
shows a useful scale-invariance: for a given stellar temperature ($T_\star$) and chemical composition, the EW of the emission lines depend in first approximation only on the volume emission measure of the wind normalized to the area of
the stellar surface. An equivalent quantity, which has been introduced by \citet{Schmutz1989}, is the transformed radius $R_\mathrm{t}$ defined as:
\begin{equation}
    R_\mathrm{t} = R_{\star} \left[\frac{\text{v}_{\infty}}{2500~\mathrm{km}~\mathrm{s}^{-1}} \bigg/ \frac{\dot{M} \sqrt{D}}{10^{-4} \mathrm{M}_{\odot}~\mathrm{yr}^{-1}}\right]^{2/3}~.  
\label{eq:r_t}
\end{equation}
Different combinations of stellar radii ($R_\star$) and mass-loss rates ($\dot{M}$)
can thus lead to the same emission-line strengths. This invariance also includes the micro-clumping
parameter $D$, which is defined as the density contrast between wind clumps and a smooth wind of the same $\dot{M}$. 
Hence, empirically derived $\dot{M}$ depend on the adopted value of $D$.
The parameters of our best fitting model for the stellar atmosphere of HD\,193949 obtained with PoWR are listed in Table~\ref{tab:powrparameters}.
All of our calculations were performed adopting a stellar mass of $M_{\star}$=0.6~M$_{\odot}$, which is the typical value for an evolved low-mass star \citep[see][and references therein]{MillerBertolami2016,Kepler2016}. The actual mass has only little impact on the spectrum of the central star, as the spectrum is formed in the wind.  
The stellar temperature $T_\star$, defined at $\tau_\mathrm{Ross}$=20, is constrained by the relative strength of the emission lines, mainly by the ratio of the \ovi\ vs.\ the \ov\ emission lines. The best fit is achieved with $T_\star=140^{\,+5}_{\,-2}$\,kK. 

\begin{figure*}
\begin{center}
\includegraphics[angle=0,width=0.9\linewidth]{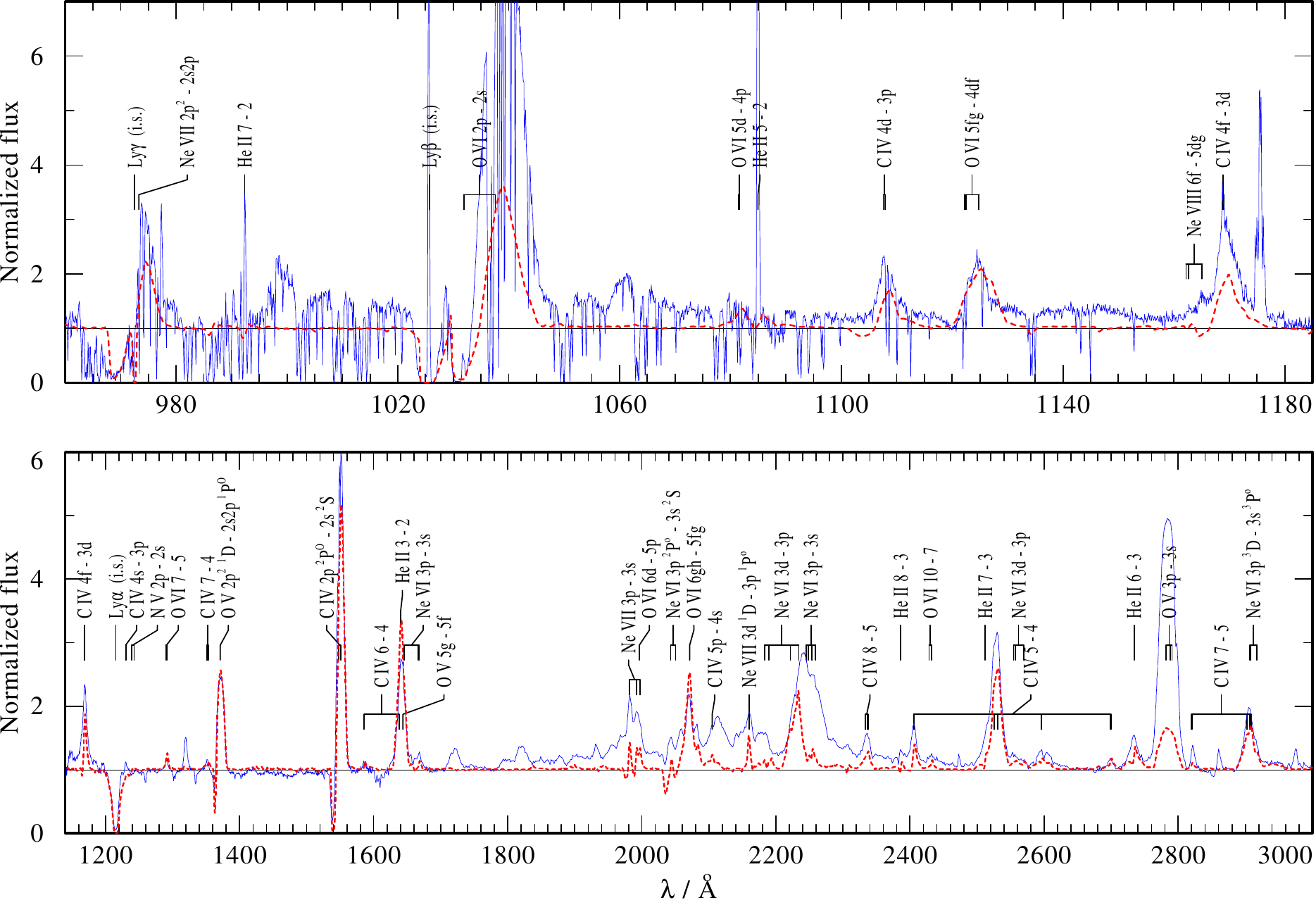}
\includegraphics[angle=0,width=0.9\linewidth]{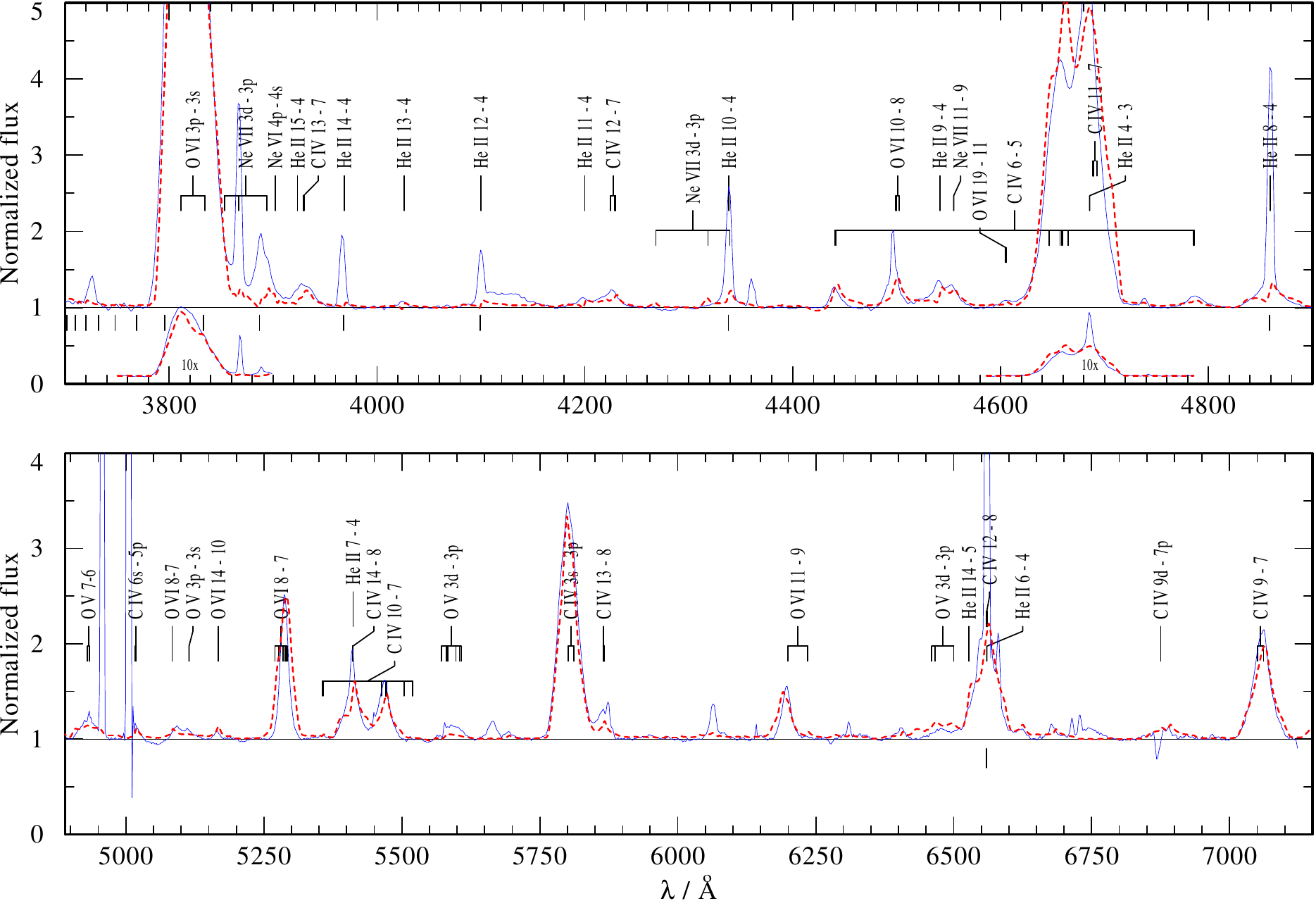}
\caption{PoWR model (red dashed line) compared with the UV and optical
observations of the CSPN HD\,193949 (blue): \emph{FUSE} (1st panel), \emph{HST} STIS (2nd panel)
and optical ALFOSC (3rd and 4th panel). The model spectra are convolved with a Gaussian to match the instrument's spectral resolution. The black lines below the continuum in the 3rd and 4th panel mark the nebular Balmer lines. See text for details.}
\label{fig:powr_uv}
\end{center}
\end{figure*}

\begin{table}
  \caption{Parameters of HD\,193949, the CSPN of NGC\,6905, obtained with PoWR.}
\setlength{\tabcolsep}{0.7\tabcolsep}
  \label{tab:powrparameters}
\begin{tabular}{lcl}
\hline
Parameter & Value & Comment\\
\hline
$d$ [kpc]                            & $2.7\pm0.2$       & \cite{Bailer-Jones2021} \\
$E(B-V)$ [mag]                       & $0.20\pm0.05$     & SED fit \\
$M_{\star}$ [M$_\odot$]              & 0.6               & adopted \\
$T_\star$ [kK]                       & $140^{+5}_{-2}$   & Defined at $\tau_\text{Ross}=20$ \\
$\log(L_{\star}$/L$_\odot)$            & 3.9$\pm0.1$       &  \\
$R_{\star}$ [R$_\odot$]              & $0.15\pm0.02$     & stellar radius \\ 
$\log R_\text{t}$ [R$_\odot$]        & $0.75\pm0.05$     & transformed radius (see Eq.~3)\\
$\dot{M}/$M$_{\odot}$~yr$^{-1}$      & $(1.1\pm0.3)\times10^{-7}$            &  \\ 
$\text{v}_{\infty}$ [km\,s$^{-1}$]          & $2000\pm100$      &  \\
$D$                                  & 10                & adopted density contrast \\
$\beta$                              & 1                 & $\beta$-law exponent\\
\hline
\multicolumn{3}{l}{Chemical abundances (mass fraction)}\\
\hline
H  &  $<0.05$          &  upper limit\\
He &  $0.55\pm0.05$    & \\
C  &  $0.35\pm0.05$    & \\
N  &  $<6.9\times10^{-5}$  & $1/10$ solar, upper limit \\
O  &  $0.08\pm0.01$    & \\
Ne &  $0.02\pm0.01$    & \\
Fe &  $1.4\times10^{-3}$       & solar, adopted \\
\hline
\end{tabular}
\end{table}

With $T_\star$ and $R_\mathrm{t}$ determined from the continuum-normalized spectra, we can fit the synthetic SED to {\it Gaia} eDR3 photometry \citep{Gaia2020} and the flux-calibrated UV spectra to estimate $L_\star$ and $E(B-V)$. Taking into account the loss of flux from the CSPN due to the long-slit spectroscopic observations, the stellar luminosity is $L_{\star}$=7600~L$_{\odot}$ for the photogeometric distance of 2.7~kpc \citep{Bailer-Jones2021}. Our fit implies a reddening of $E(B−V) = 0.21$\,mag using the extinction law by \citet{Cardelli1989}, consistent with value derived by previous authors \citep[see][]{2014Keller}. We note, however, that such model requires a higher value of $R_\mathrm{V}$=3.8 that suggests that the extinction towards HD\,193949 is higher due to the presence of different dust compositions along the line of sight \citep[see, e.g.,][]{Todt2010}.

The spectral fit to the blue edges of the P-Cygni line profiles resulted in a terminal wind velocity of $\text{v}_{\infty}=2000$~km~s$^{-1}$ by adopting a $\beta$-law with $\beta=1$. Additional line broadening by microturbulence is also included in our models. From the shape of the line profiles we estimate a value of about $100\,\mathrm{km}\,\mathrm{s}^{-1}$. From Equation~(\ref{eq:r_t}) we obtained a mass-loss rate of $\dot{M}=1.1\times10^{-7}$~M$_\odot$~yr$^{-1}$ adopting a density contrast of $D$=10, which is a typical value also found for other [WC]-CSPNe \citep[e.g.,][]{Marcolino2007,Todt2008}. We achieve a similar fit quality also with a lower value of $D=4$ and $\log(R_\mathrm{t}/R_\odot)=0.8$, while even smaller values of $D$, e.g., a smooth wind with $D=1$, yield too strong electron scattering line wings. For larger values of $D$ the fit quality is worse.

To determine the C and He abundances we mainly focused on  the "diagnostic line pair" He{\,\scshape ii}\,$\lambda$5411 \& C{\,\scshape iv}\,$\lambda$5471 \citep{Todt2006}. Both features are formed at the same region in the wind, hence their relative strength is less sensitive to temperature and $\dot{M}$. For the hot [WC] and [WO] stars an abundance ratio of about He:C $\approx$~60:30 yields approximately equal strength for both emission features, as it is observed.   
For the O abundance we used primarily the O{\,\scshape vi} multiplet at $\approx$~5290~\AA, as this line is less sensitive to $T_\star$ and $R_\mathrm{t}$ and more sensitive to the O abundance. 
The spectrum of HD\,193949 displays many Ne{\,\scshape vi} and Ne{\,\scshape vii} lines in the UV and optical. With the inferred Ne abundance of about 2 per cent by mass we got a reasonable fit to most of these lines. 
However, we also note the presence of \neviii\ features around 1162\,\AA{} and 1165\,\AA{} (see Fig.~\ref{fig:powr_uv}, top panel), as described in \citet{Werner2007}.
While these lines are in absorption in the model, it appears that they are in emission in the {\it FUSE} observation. There is also the emission feature at 6068\,\AA{} in the NOT spectrum, which might be also attributed to \neviii. However, our model is not hot enough
to produce this potential \neviii\ emission line. For our object, this would require a stellar temperature above 170\,kK, which is excluded by the presence of the observed O{\,\scshape v} lines. 
This emission feature has been reported to appear in the spectrum of other hot CSPNe such as NGC\,2371 (see Paper~I) and PC\,22 (Sabin et al. in prep.).
It seems that there is no N line present in the UV and optical spectra: the N{\,\scshape v} resonance doublet is not detectable and the emission features around 4604~\AA\ and 4933~\AA, which are also seen in hot WN stars, originate from O{\,\scshape vi} and O{\,\scshape v} respectively. Therefore, we estimate the N fraction to be at most 1/10 of the solar value.   
For the H abundance we can only give an upper limit too, as stellar Balmer lines would be blended with the corresponding He{\,\scshape ii} lines from the Pickering series and nebular H emission lines. We find that a H mass fraction below 5~per cent would escape detection. 
In the absence of strong spectral lines from the iron group elements and with regard to the limited quality of the UV spectra we could not determine an iron abundance and adopted instead the solar value.  
Figure~\ref{fig:powr_uv} shows the comparison between the best fitting PoWR model for HD\,193949
and its \emph{FUSE}, {\it HST} STIS and NOT optical spectra, respectively.

Our analysis yields similar values for the abundances and stellar parameters reported in previous works \citep{Marcolino2007,2014Keller}, although this work is based on a measured distance, hence we can give now a reliable estimate for $L$ and $\dot{M}$. 
We note that our C abundance is lower than that derived 
by \citet{2014Keller}, who inferred the abundances from the UV spectra only, while we fitted also the optical spectrum. 

\subsection{Physical properties of NGC\,6905}
\label{sec:physical}

\begin{figure*}
\begin{center}
\includegraphics[angle=0,width=1.0\linewidth]{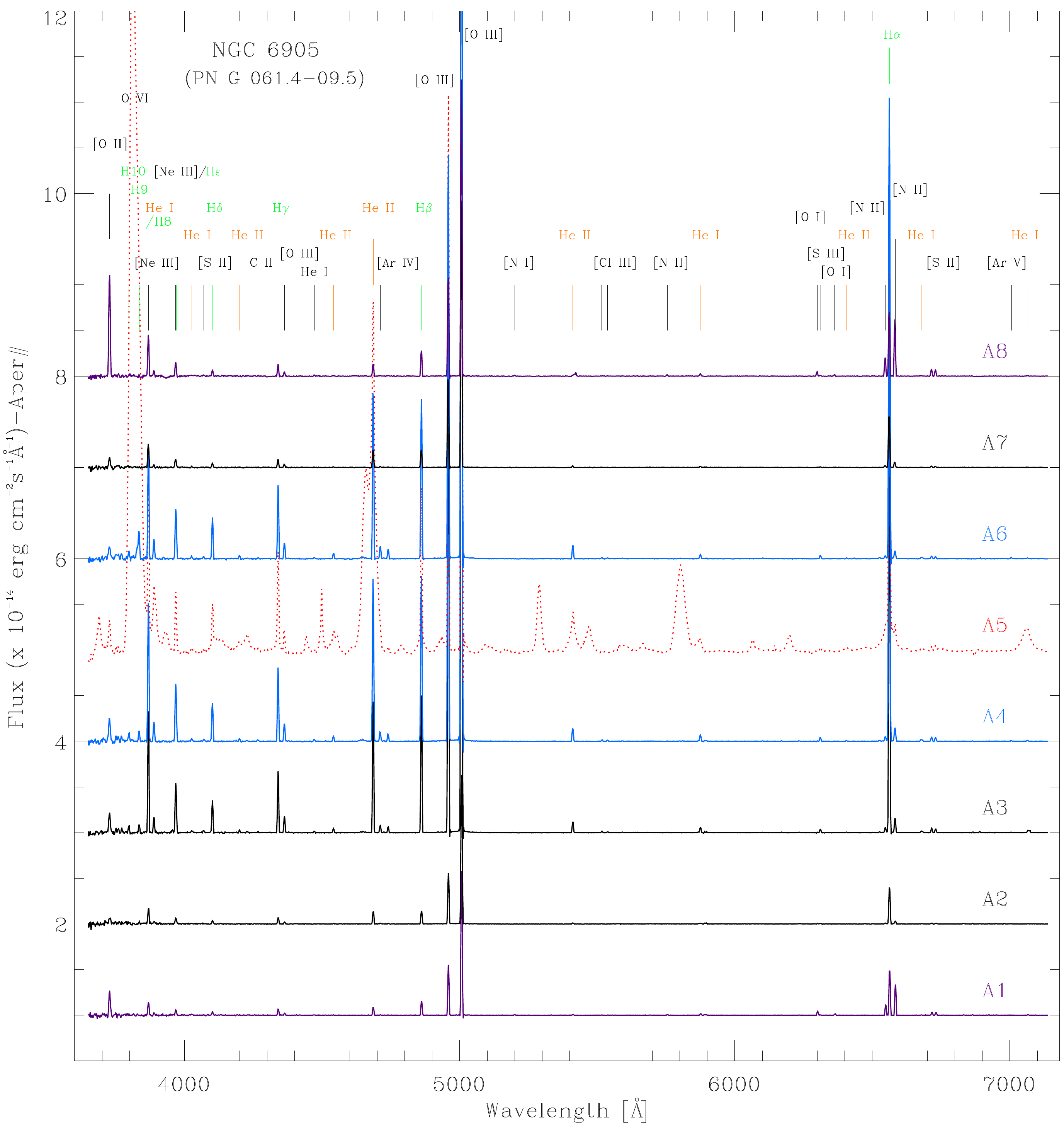}
\caption{NOT ALFOSC spectra from the extracted regions A1--A8 in NGC\,6905.
The main emission lines are labeled. For the sake of clarity, H-Balmer lines are indicated in green, He lines in orange and the rest of the lines in black.
The spectrum of the CSPN HD\,19394 (A5) is shown in red dotted line in the middle for reference.
The colours of the spectra were coded as the extraction apertures indicated in Fig.~\ref{fig:ngc6909_opt}.
To avoid overlap, the spectra are shifted upwards (A6--A8) and downwards (A4--A1) with respect to the spectrum of the CSPN by appropriate amounts (its continuum level corresponds to its aperture number).
Thereby, the "onion-like" ionization structure of the object can be clearly appreciated.
See Table~\ref{tab:sample} for details of the present lines and their fluxes.
}
\label{fig:spec_examples}
\end{center}
\end{figure*}

\begin{table*}
\begin{center}
\caption[]{De-reddened line fluxes relative to \hb=100. The location of the apertures A1--A8 are shown 
in Fig.~\ref{fig:ngc6909_opt} and their sizes are indicated. Their spectra are shown in Fig.~\ref{fig:spec_examples}.}
\setlength{\tabcolsep}{0.6\tabcolsep}
\begin{tabular}{cccccccccc}
\hline
\multicolumn{1}{c}{$\lambda_{0}$}&
\multicolumn{1}{c}{line}&
\multicolumn{1}{c}{A1}&
\multicolumn{1}{c}{A2}&
\multicolumn{1}{c}{A3}&
\multicolumn{1}{c}{A4}&
\multicolumn{1}{c}{A6}&
\multicolumn{1}{c}{A7}&
\multicolumn{1}{c}{A8}\\
\multicolumn{1}{c}{}&
\multicolumn{1}{c}{}&
\multicolumn{1}{c}{(5 arcsec)}&
\multicolumn{1}{c}{(5 arcsec)}&
\multicolumn{1}{c}{(5 arcsec)}&
\multicolumn{1}{c}{(5 arcsec)}&
\multicolumn{1}{c}{(5 arcsec)}&
\multicolumn{1}{c}{(5 arcsec)}&
\multicolumn{1}{c}{(5 arcsec)}\\
\hline
3727.0 & \oii\   & 222$\pm$12  &  \dots           & 21.1$\pm$1.1 &  18.8$\pm$3.9 &   9.8$\pm$2.9 &   64.0$\pm$3.4 &   490$\pm$27  \\
3797.9 &  H10    &  \dots         &  \dots           &  5.4$\pm$0.3 &   5.4$\pm$1.5 &   4.0$\pm$1.9 & \dots             &  \dots          \\
3835.4 &  H9     &  \dots         &  \dots           &  6.4$\pm$0.3 &   6.3$\pm$1.3 &  20.3$\pm$2.2 & \dots             &  \dots          \\
3868.8 & \neiii\ & 108$\pm$6   & 113$\pm$6     & 94.3$\pm$4.7 &  90$\pm$6 &  70.4$\pm$4.3 &    134$\pm$7   &   170$\pm$11 \\
3888.6 & \hei\   &18.9$\pm$2.0 &16.7$\pm$3.0   & 12.5$\pm$0.6 &  12.6$\pm$1.5 &  12.0$\pm$1.4 &   11.8$\pm$0.8 &   20.5$\pm$3.8 \\
3967.5 & \neiii\ &45.6$\pm$2.3 &46.3$\pm$2.3   & 42.9$\pm$4.9 &  40.1$\pm$2.5 &  35.0$\pm$2.2 &   54.1$\pm$2.9 &   56$\pm$4 \\
4026.2 & \hei\   &  \dots         &  \dots           &  1.9$\pm$0.5 &   1.6$\pm$0.3 &   1.3$\pm$0.3 & \dots             &  \dots           \\   
4070.0 & \sii\   &  \dots         &  \dots           &  1.9$\pm$0.4 &   1.5$\pm$0.3 &   1.3$\pm$0.3 &    \dots          &    4.9$\pm$1.7 \\   
4101.7 & \hd\    &25.7$\pm$1.3 &   25$\pm$10   & 24.8$\pm$1.6 &  25.3$\pm$1.4 &  25.9$\pm$1.4 &   25.1$\pm$1.3 &   25.0$\pm$1.9 \\   
4200.0 & \heii\  &  \dots         &  \dots           &  1.7$\pm$0.3 &   1.6$\pm$0.2 &   2.0$\pm$0.3 &    3.0$\pm$0.4 &  \dots           \\   
4267.3 & \cii\   &  \dots         &  \dots           & \dots           &   1.1$\pm$0.2 &   0.6$\pm$0.2 & \dots             &  \dots           \\
4340.5 & \hg\    &46.2$\pm$3.6 & 47.6$\pm$3.3  & 45.8$\pm$2.3 &  45.8$\pm$2.3 &  45.7$\pm$2.3 &   47$\pm$5 &   45.8$\pm$2.4 \\   
4363.2 & \oiii\  &13.8$\pm$2.1 & 14.4$\pm$1.9  & 12.4$\pm$0.7 &  11.0$\pm$0.6 &  10.2$\pm$0.6 &   18.3$\pm$2.9 &   16.8$\pm$1.2 \\   
4471.5 & \hei\   & 2.7$\pm$1.5 &  \dots           &  1.0$\pm$0.2 &   1.2$\pm$0.2 &   0.7$\pm$0.1 &  \dots            &    3.9$\pm$0.8 \\   
4541.0 & \heii\  &  \dots         &  \dots           &  3.0$\pm$0.3 &   3.0$\pm$0.2 &   3.3$\pm$0.3 &    2.7$\pm$1.0 &    2.3$\pm$1.0 \\   
4686.0 & \heii\  &56.8$\pm$3.0 & 96$\pm$5  & 99$\pm$5 &  100$\pm$5 &   106$\pm$5   &   97$\pm$5 &   48.0$\pm$2.5 \\   
4711.4 & \ariv\  &  \dots         &  4.8$\pm$1.3  &  5.4$\pm$0.3 &   5.8$\pm$0.3 &   7.7$\pm$0.4 &    4.3$\pm$1.2 &    1.9$\pm$0.8 \\   
4740.2 & \ariv\  &  \dots         &  \dots           &  4.1$\pm$0.3 &   4.3$\pm$0.3 &   5.8$\pm$0.3 &    2.4$\pm$1.1 &    2.2$\pm$0.9 \\   
4861.4 & \hb\    & 100            &  100             & 100             &  100             & 100              & 100               & 100                \\   
4958.9 & \oiii\  &  347$\pm$18 &  385$\pm$20   & 353$\pm$18   &   310$\pm$16  &   249$\pm$13  &    507$\pm$27  &    391$\pm$20 \\   
5006.8 & \oiii\  &1040$\pm$60  & 1170$\pm$60   &1050$\pm$50   &   930$\pm$50  &   743$\pm$38  &   1510$\pm$90  & 1160$\pm$60\\  
5200.0 & \nia\   & 3.6$\pm$0.9 &   \dots          & \dots           &  \dots           &  \dots           &  \dots            &    2.4$\pm$0.6 \\
5411.0 & \heii\  & 4.7$\pm$1.0 &  7.1$\pm$1.0  & 7.6$\pm$0.4  &   7.6$\pm$0.4 &   8.1$\pm$0.4 &    7.6$\pm$0.7 & \dots              \\  
5517.7 & \cliii\ & 1.5$\pm$0.9 &  \dots           & 0.9$\pm$0.1  &   0.7$\pm$0.1 &   0.7$\pm$0.1 &    1.3$\pm$0.7 & \dots \\
5537.9 & \cliii\ & 1.1$\pm$0.9 &  \dots           & 0.6$\pm$0.1  &   0.6$\pm$0.1 &   0.5$\pm$0.1 &  \dots            & \dots              \\   
5754.6 & \nii\   & 3.9$\pm$0.9 &  \dots           & 0.2$\pm$0.1  & \dots            &  \dots           &  \dots            &   4.6$\pm$0.5 \\   
5875.6 & \hei\   & 9.9$\pm$1.1 &  6.2$\pm$1.4  & 3.7$\pm$0.2  &   3.6$\pm$0.2 &   2.5$\pm$0.2 &    5.8$\pm$1.0 &   9.3$\pm$0.7 \\ 
6300.0 & \oi\    &23.6$\pm$1.7 &  \dots           & \dots           &  \dots           &  \dots           &  \dots            &  15.3$\pm$1.0 \\
6312.1 & \siii\  & 2.3$\pm$1.4 &  2.8$\pm$1.5  & 2.2$\pm$0.2  &   2.1$\pm$0.2 &   1.9$\pm$0.2 &    2.8$\pm$1.1 &   2.8$\pm$0.7 \\   
6363.8 & \oi\    & 8.1$\pm$1.3 &  \dots           & \dots           &  \dots           &  \dots           &  \dots            &   5.0$\pm$0.7 \\   
6406.0 & \heii\  & \dots          &  \dots           & 0.3$\pm$0.1  &   0.3$\pm$0.1 &   0.4$\pm$0.1 &  \dots            & \dots              \\   
6548.1 & \nii\   &60.6$\pm$3.1 &  7.1$\pm$1.2  & 3.4$\pm$0.2  &   2.7$\pm$0.2 &   1.9$\pm$0.2 &    9.6$\pm$1.1 &  67.9$\pm$3.4 \\   
6562.8 & \ha\    & 283$\pm$15  &  285$\pm$15   & 260$\pm$13   &   271$\pm$14  & 279$\pm$14    &     278$\pm$15 & 245$\pm$12 \\   
6583.5 & \nii\   & 187$\pm$10  & 21.9$\pm$1.5  & 9.8$\pm$0.5  &   7.8$\pm$0.4 &   5.0$\pm$0.3 &   27.6$\pm$1.5 & 208$\pm$10 \\
6678.2 & \hei\   & 1.7$\pm$0.9 &  1.7$\pm$1.1  & 1.4$\pm$0.1  &   1.4$\pm$0.1 & \dots            &    1.4$\pm$0.8 &   2.0$\pm$0.5 \\
6716.4 & \sii\   &17.4$\pm$1.2 &  5.4$\pm$1.1  & 2.7$\pm$0.2  &   2.2$\pm$0.1 &   1.5$\pm$0.1 &    7.0$\pm$0.9 &  24.8$\pm$1.3 \\
6730.8 & \sii\   &15.5$\pm$1.1 &  4.2$\pm$1.1  & 2.4$\pm$0.2  &   2.0$\pm$0.1 &   1.2$\pm$0.1 &    5.1$\pm$0.8 &  22.9$\pm$1.2 \\
7006.0 & \arv\   & \dots          &  \dots           &   \dots         &   0.4$\pm$0.1 &   0.6$\pm$0.1 & \dots             & \dots              \\
7065.2 & \hei\   & 2.4$\pm$1.6 &  \dots           &   \dots         &   0.5$\pm$0.2 &   0.4$\pm$0.1 & \dots             &   2.0$\pm$0.9 \\
\hline
log(H$\beta$) &[erg~cm$^{-2}$~s$^{-1}$] &  $-$13.82$\pm$0.03 &  $-$14.03$\pm$0.02 & $-$12.89$\pm$0.01 &  $-$12.79$\pm$0.01 &  $-$12.89$\pm$0.01 & $-$13.79$\pm$0.03 & $-$13.69$\pm$0.01\\ 
$c$(H$\beta$) &                      & 0.23$\pm$0.02           & 0.05$\pm$0.04            &  0.20$\pm$0.09         &  0.19$\pm$0.06         & 0.10$\pm$0.05           &  0.15$\pm$0.04          &0.11$\pm$0.08 \\
$A_{\rm V}$   & [mag]                   & 0.49$\pm$0.04           & 0.10$\pm$0.09            &  0.41$\pm$0.20         &  0.39$\pm$0.12         & 0.22$\pm$0.10           &  0.32$\pm$0.07          & 0.24$\pm$0.17 \\
\hline
\end{tabular}
\vspace{0.4cm}
\label{tab:sample}
\end{center}
\end{table*}

\begin{table*}
\centering
\small\addtolength{\tabcolsep}{-4pt}
\begin{center}
\caption[]{Mean and standard deviation ($\sigma$) of the MC distribution for the $T_\mathrm{e}$ and $n_\mathrm{e}$ as well as the ionic and total abundances for each region of NGC\,6905.}
\setlength{\tabcolsep}{0.3\tabcolsep}
\begin{adjustbox}{width=18cm,center}
\begin{tabular}{@{\extracolsep{4pt}}llccccccc}
\hline
Parameter & Condition & A1 & A2 & A3 & A4 & A6 & A7 & A8\\
\hline
$T_\mathrm{e}$(\nii)~[K]  & $n_\mathrm{e}$(\sii)          & 11470$\pm$1350 & \dots           &10660$\pm$2180 &\dots            & \dots           & \dots           &11560$\pm$690 \\
$T_\mathrm{e}$(\nii)~[K] &$n_\mathrm{e}$(\ariv)     & \dots          & \dots        &10610$\pm$2180 &\dots         & \dots        & \dots        &10350$\pm$1290 \\
$T_\mathrm{e}$(\oiii)~[K] & $n_\mathrm{e}$(\sii)          & 12840$\pm$1010 &12430$\pm$800 &12030$\pm$380  &12190$\pm$380 &12930$\pm$440 &12310$\pm$960 &12800$\pm$520 \\
$T_\mathrm{e}$(\oiii)~[K] & $n_\mathrm{e}$(\cliii)        & \dots             & \dots           & \dots            &12160$\pm$380 &12920$\pm$440 & \dots           & \dots         \\
$T_\mathrm{e}$(\oiii)~[K] & $n_\mathrm{e}$(\ariv)         & \dots             & \dots           &12010$\pm$380  &12170$\pm$380 &12910$\pm$440 & \dots           &12500$\pm$720 \\
\hline
$n_\mathrm{e}$(\sii)~[cm$^{-3}$]   & $T_\mathrm{e}$(\nii)  & 500$\pm$290   & \dots           &480$\pm$220    & \dots           & \dots           & \dots           &560$\pm$210 \\
$n_\mathrm{e}$(\sii)~[cm$^{-3}$]   & $T_\mathrm{e}$(\oiii) & 520$\pm$300   &\dots         &510$\pm$230    &560$\pm$230   &360$\pm$250   &400$\pm$470   &580$\pm$220 \\
$n_\mathrm{e}$(\cliii)~[cm$^{-3}$] & $T_\mathrm{e}$(\oiii) & \dots            & \dots           & \dots            &1930$\pm$1370 &1130$\pm$960  & \dots           & \dots       \\ 
$n_\mathrm{e}$(\ariv)~[cm$^{-3}$]  & $T_\mathrm{e}$(\oiii) & \dots            & \dots           &960$\pm$670    &820$\pm$610   &840$\pm$610   & \dots           & \dots       \\
$n_\mathrm{e}$(\ariv)~[cm$^{-3}$] &$T_\mathrm{e}$(\nii) &\dots       &\dots         &990$\pm$710    &\dots         &\dots         &\dots         &\dots     \\

\hline
Ionic abundances \\
log(Ar$^{+3}$/H$^{+}$)&&  \dots                         &(6.4$\pm$2.3)$\times10^{-7}$ &(7.6$\pm$0.8)$\times10^{-7}$ &(7.8$\pm$0.8)$\times10^{-7}$ &(8.8$\pm$0.9)$\times10^{-7}$ &(5.9$\pm$2.4)$\times10^{-7}$ &(2.9$\pm$0.9)$\times10^{-7}$ \\
log(Ar$^{+4}$/H$^{+}$)&&  \dots                         & \dots                          & \dots                          &(4.8$\pm$1.4)$\times10^{-8}$ &(7.0$\pm$1.7)$\times10^{-8}$ & \dots                          & \dots                          \\ 
log(Cl$^{+2}$/H$^{+}$)&&  \dots                         & \dots                          &1.6$\times10^{-7}$:          &(4.8$\pm$0.7)$\times10^{-8}$ &(3.5$\pm$0.5)$\times10^{-8}$ & \dots                          & \dots                          \\
log(He$^{+}$/H$^{+}$) &&(9.1$\pm$1.2)$\times10^{-2}$ &(8.8$\pm$1.7)$\times10^{-2}$ &(5.2$\pm$0.6)$\times10^{-2}$ &(4.5$\pm$0.5)$\times10^{-2}$ &(3.9$\pm$0.5)$\times10^{-2}$ &(6.9$\pm$0.9)$\times10^{-2}$ &(10.3$\pm$1.8)$\times10^{-2}$\\
log(He$^{+2}$/H$^{+}$)&&(5.0$\pm$0.7)$\times10^{-2}$ &(7.9$\pm$0.8)$\times10^{-2}$ &(7.6$\pm$0.7)$\times10^{-2}$ &(7.7$\pm$0.6)$\times10^{-2}$ &(8.7$\pm$0.7)$\times10^{-2}$ &(9.4$\pm$1.2)$\times10^{-2}$ &(4.8$\pm$1.4)$\times10^{-2}$ \\
log(N$^{0}$/H$^{+}$)  &&(3.9$\pm$3.0)$\times10^{-6}$ & \dots                          & \dots                          &  \dots                         & \dots                          & \dots                          &(2.4$\pm$0.9)$\times10^{-6}$ \\
log(N$^{+}$/H$^{+}$)  &&(2.9$\pm$1.1)$\times10^{-5}$ &(2.7$\pm$0.5)$\times10^{-6}$ &3.0$\times10^{-6}$:          &(1.1$\pm$0.1)$\times10^{-6}$ &(6.1$\pm$0.7)$\times10^{-7}$ &(3.7$\pm$0.8)$\times10^{-6}$ &(3.5$\pm$0.6)$\times10^{-5}$ \\
log(Ne$^{+2}$/H$^{+}$)&&(6.1$\pm$1.8)$\times10^{-5}$ &(6.7$\pm$1.6)$\times10^{-5}$ &(6.1$\pm$0.8)$\times10^{-5}$ &(5.7$\pm$0.7)$\times10^{-5}$ &(4.0$\pm$0.5)$\times10^{-5}$ &(7.0$\pm$2.2)$\times10^{-5}$ &(6.9$\pm$1.0)$\times10^{-5}$ \\
log(O$^{0}$/H$^{+}$)  &&(3.3$\pm$2.1)$\times10^{-5}$ & \dots                          & \dots                          &  \dots                         & \dots                          & \dots                          &(2.1$\pm$0.5)$\times10^{-5}$ \\
log(O$^{+}$/H$^{+}$)  &&(1.3$\pm$1.2)$\times10^{-4}$ & \dots                          &\dots                        &(6.7$\pm$1.8)$\times10^{-6}$ &(3.0$\pm$1.0)$\times10^{-6}$ &(2.4$\pm$0.8)$\times10^{-5}$ &(1.9$\pm$0.5)$\times10^{-4}$ \\
log(O$^{+2}$/H$^{+}$) &&(1.7$\pm$0.4)$\times10^{-4}$ &(2.1$\pm$0.4)$\times10^{-4}$ &(2.1$\pm$0.2)$\times10^{-4}$ &(1.8$\pm$0.2)$\times10^{-4}$ &(1.2$\pm$0.1)$\times10^{-4}$ &(2.9$\pm$0.7)$\times10^{-4}$ &(1.9$\pm$0.3)$\times10^{-4}$ \\
log(S$^{+}$/H$^{+}$)  &&(6.9$\pm$2.4)$\times10^{-7}$ &(1.7$\pm$0.6)$\times10^{-7}$ &\dots                        &(1.8$\pm$0.3)$\times10^{-7}$ &(1.3$\pm$0.3)$\times10^{-7}$ &(2.1$\pm$0.5)$\times10^{-7}$ &(1.2$\pm$0.3)$\times10^{-6}$ \\
log(S$^{+2}$/H$^{+}$) &&  \dots                         & \dots                          &\dots                        &(2.2$\pm$0.4)$\times10^{-6}$ &(1.7$\pm$0.3)$\times10^{-6}$ &(3.1$\pm$1.6)$\times10^{-6}$ &(4.1$\pm$1.5)$\times10^{-6}$ \\
\hline
Total abundances\\
He &$^{\dagger}$ 0.084            &0.141$\pm$0.015              &0.168$\pm$0.021              &0.128$\pm$0.011              &0.121$\pm$0.009              &0.126$\pm$0.010             &0.163$\pm$0.018              &0.152$\pm$0.023                \\ 
O  &$^{\dagger}$5.4$\times10^{-4}$&(3.9$\pm$1.6)$\times10^{-4}$ &(3.3$\pm$0.7)$\times10^{-4}$ &(4.4$\pm$1.4)$\times10^{-4}$ &(3.7$\pm$0.5)$\times10^{-4}$ &(2.9$\pm$0.4)$\times10^{-4}$&(5.7$\pm$1.5)$\times10^{-4}$ &(4.9$\pm$0.8)$\times10^{-4}$   \\ 
N  &$^{\dagger}$7.2$\times10^{-5}$&(8.7$\pm$1.9)$\times10^{-5}$ & \dots                          &(5.2$\pm$1.6)$\times10^{-5}$ &(4.9$\pm$2.0)$\times10^{-5}$ &5.7$\times10^{-5}$:         &(6.8$\pm$1.7)$\times10^{-5}$ &(8.2$\pm$1.3)$\times10^{-5}$            \\ 
Ne &$^{\dagger}$1.1$\times10^{-4}$&(8.1$\pm$2.4)$\times10^{-5}$ &(1.0$\pm$0.3)$\times10^{-4}$ &(1.2$\pm$0.2)$\times10^{-4}$ &(1.2$\pm$0.2)$\times10^{-4}$ &(1.0$\pm$0.2)$\times10^{-4}$&(1.3$\pm$0.4)$\times10^{-4}$ &(8.9$\pm$1.5)$\times10^{-5}$ \\ 
S  &$^{\dagger}$1.5$\times10^{-5}$&(3.1$\pm$0.7)$\times10^{-6}$ & \dots                          &(1.2$\pm$0.3)$\times10^{-5}$ &(1.0$\pm$0.4)$\times10^{-5}$ &1.4$\times10^{-5}$:         &(5.1$\pm$1.3)$\times10^{-6}$ &(4.3$\pm$0.8)$\times10^{-6}$           \\ 
Cl &$^{\dagger}$1.8$\times10^{-7}$& \dots                          & \dots                          &3.8$\times10^{-7}$:          &(2.1$\pm$0.4)$\times10^{-7}$ &(2.0$\pm$0.4)$\times10^{-7}$&  \dots                         & \dots                          \\ 
\hline
He/He$_\odot$            & \dots  & 1.68                         & 2.0                       & 1.52                         & 1.45                        & 1.50                       & 1.94                        & 1.81 \\
N/O          & $^{\dagger}$0.13   & 0.22$\pm$0.04                & \dots                        & 0.181$\pm$0.001              & 0.13$\pm$0.04               & 0.197:                     & 0.119$\pm$0.002             & 0.167$\pm$0.001 \\
Ne/O         & $^{\dagger}$0.21   & 0.21$\pm$0.02                & 0.30$\pm$0.03             & 0.27$\pm$0.04                & 0.32$\pm$0.01               & 0.34$\pm$0.02              & 0.23$\pm$0.01               & 0.182$\pm$0.001 \\
S/O& $^{\dagger}$3$\times10^{-2}$ & (0.8$\pm$0.2)$\times10^{-2}$ & \dots                        & (2.7$\pm$0.2)$\times10^{-2}$ & (2.6$\pm$0.7)$\times10^{-2}$& 4.8$\times10^{-2}$:        & (9.0$\pm$0.1)$\times10^{-3}$& (8.8$\pm$0.2)$\times10^{-3}$ \\ 
Cl/O         & $^{\dagger}$3$\times10^{-4}$ & \dots                 & \dots                        & 8.6$\times10^{-4}$:          &(5.5$\pm$0.3)$\times10^{-4}$ &(6.9$\pm$0.4)$\times10^{-4}$& \dots                       & \dots \\
\hline
\end{tabular}
\end{adjustbox}
$^{\dagger}$Solar values taken from \citet[][]{Lodders2010}
\label{tab:sample2}
\end{center}
\end{table*}

The seven ID spectra obtained from regions A1--A4 and A6--A8 (see Figure~\ref{fig:ngc6909_opt}) corresponding to the nebular emission of NGC\,6905 are shown in Figure~\ref{fig:spec_examples}. Fluxes of the nebular lines were measured using the Gaussian line fitting option of the splot task in {\scshape iraf}. The resultant line fluxes and statistical errors are listed in Table~\ref{tab:sample}.
Following Sabin et al. 2021 in prep., we quadratically add 5~per cent to the statistical errors described in Section~3.1  to  take  into  account plausible additional uncertainties  from the various calibrations, the reddening correction and the uncertainties on the atomic data used to derive the abundances.

The spectra have been analyzed by means of the {\scshape PyNeb} code \citep[version 1.1.15;][]{Luridiana2015}. First, the correction from extinction was performed using the \citet{Cardelli1989} law with $R_\mathrm{V}$=3.1. The logarithmic extinction $c$(H$\beta$) was found to be in the range 0--0.22, consistent with that estimated from UV analysis towards the CSPN (see Table~\ref{tab:powrparameters}).
Our NOT ALFOSC spectra are the highest-quality spectra of NGC\,6905 presented so far \citep[see, e.g.,][]{Kingsburgh1994,Pena1998}, which will allow us to assess possible physical and/or abundance differences within this WRPN.

We applied a Monte Carlo (MC) procedure to determine the propagation of the uncertainties from the emission lines into the subsequent determination of physical parameters and abundances. The details are described in length in Sabin et al.\,(in prep.) for the case of the WRPN PC\,22. For all the parameters derived in the following we present the mean and standard deviation obtained from the MC analysis. We did not consider in our analysis the lines showing an error on their measurements greater than 50 per cent. 
The electron temperature ($T_\mathrm{e}$) and density ($n_\mathrm{e}$) are calculated for all the available ions and we distinguish the low and high ionization potentials with the \nii\ and \sii\ in one hand, and the \oiii, \cliii\ and \ariv\ on the other hand.
The estimated $T_\mathrm{e}$ and $n_\mathrm{e}$ values for the different regions in NGC\,6905 are listed in Table~\ref{tab:sample2}. We note for example that three different estimates for the $T_\mathrm{e}$(\oiii) were attempted which correspond to three $n_\mathrm{e}$ estimations using the \sii, \cliii\ and the \ariv\ doublets. Conversely, the corresponding $n_\mathrm{e}$(\sii), $n_\mathrm{e}$(\cliii) and $n_\mathrm{e}$(\ariv) were computed using the $T_\mathrm{e}$(\oiii).

The ionic and total elemental abundances were derived using the ionisation correction factors (ICFs) derived by \citet{DIMS2014} adopting the values of $T_\mathrm{e}$([O\,{\sc iii}]) and $n_\mathrm{e}$([S\,{\sc ii}]) for each region.
For some elements, different values for the total abundances can be estimated based on the equations used for the ICFs, the selection is made based on the values {$\upsilon$} = He$^{++}$/(He$^{+}$+He$^{++}$) and {$\omega$} = O$^{++}$/(O$^{+}$+O$^{++}$) and their range of validity as indicated by \citet{DIMS2014}.
Finally, in the case where the O$^{+}$ abundance is not defined (i.e. O$^{+}$/H$^{+}$= 0) we use ICF(O$^{++}$) = 1 and a similar approach was used for the He ICF.
This applied only for the region A2. Such procedure would explain the divergence (of about one order of magnitude) that we observed between the abundances found in A2 and the other regions of the nebula and we therefore discarded these results from the table.

\subsection{Photoionization model of NGC\,6905}

\begin{figure}
\begin{center}
\includegraphics[width=\linewidth]{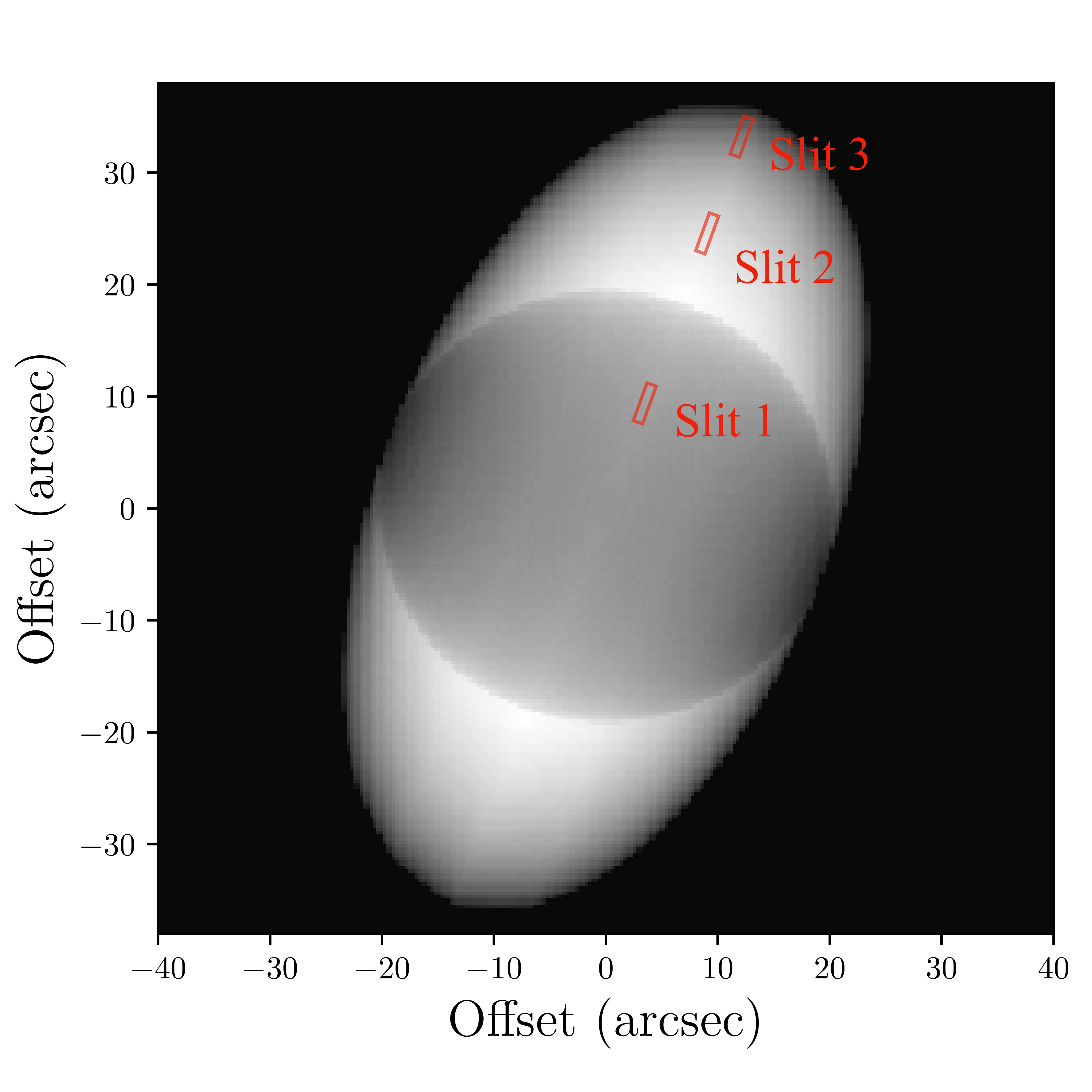}
\caption{Density distribution of} our {\scshape pyCloudy} 3D model. The rectangular regions show the relative position and apertures of the pseudo-slit for the representative regions A2, A3 and A4 in NGC\,6905. See results in Table~\ref{tab:emmision_lines}.
\label{fig:pycloudy_model}
\end{center}
\end{figure}

\begin{table*}
\begin{center}
\caption{Emission line fluxes for three representative regions (A2, A3 and A4) in NGC\,6905 (see Fig.~\ref{fig:ngc6909_opt}, left), compared to the predictions for Slit1 and Slit 2 from our {\sc Cloudy} models (see Fig.~\ref{fig:pycloudy_model}). The line fluxes are normalized with respect to \hb=100.}
\begin{tabular}{cccccccccccc}
\hline
\multicolumn{1}{c}{Line} &
\multicolumn{1}{c}{$\lambda$} &
\multicolumn{4}{c}{From observations} &
\multicolumn{3}{c}{Model A} &
\multicolumn{3}{c}{Model B}\\
\multicolumn{1}{c}{} &
\multicolumn{1}{c}{$($\AA$)$} &
\multicolumn{1}{c}{A1} &
\multicolumn{1}{c}{A2} &
\multicolumn{1}{c}{A3} &
\multicolumn{1}{c}{A4} &
\multicolumn{1}{c}{Slit 1} &
\multicolumn{1}{c}{Slit 2} &
\multicolumn{1}{c}{Slit 3} &
\multicolumn{1}{c}{Slit 1} &
\multicolumn{1}{c}{Slit 2} &
\multicolumn{1}{c}{Slit 3} \\
\hline
\oii\    & 3727 &  222$\pm$12    & \dots             &   21.1$\pm$1.1 &   18.8$\pm$3.9 &  164.8 &   198.6 &  383.7 &   142.5 &   159.3 &  200\\
\neiii\  & 3869 &  108$\pm$6     &   113$\pm$6    &   94.3$\pm$4.7 &   90.0$\pm$6.0 &  123.6 &   115.4 &  108.5 &   150.7 &   143.0 &  132\\
\neiii\  & 3968 &   45.6$\pm$2.3 &  46.3$\pm$2.3  &   42.9$\pm$4.9 &   40.1$\pm$2.5 &   37.5 &    35.0 &   32.9 &   45.7  &    43.4 &  40.1 \\
 \hd\    & 4102 &  25.7$\pm$1.3  &  25.0$\pm$10.1 &   24.8$\pm$1.6 &   25.3$\pm$1.4 &   26.2 &    26.2 &   26.2 &   26.2  &    26.2 & 26.2 \\
 \hg\    & 4341 &  46.2$\pm$3.6  &  47.6$\pm$3.3  &   45.8$\pm$2.3 &   45.8$\pm$2.3 &   47.5 &    47.4 &   47.5 &    47.7 &    47.5 &  47.5 \\
\oiii\   & 4363 &   13.8$\pm$2.1 &  14.4$\pm$1.9  &   12.4$\pm$0.7 &   11.0$\pm$0.6 &   15.2 &    13.3 &   10.9 &    22.3 &    20.5 &  17.6 \\
\hei\    & 4471 &    2.7$\pm$1.5 &  \dots            &    1.0$\pm$0.2 &    1.2$\pm$0.2 &    2.8 &     3.2 &    3.9 &     2.0 &     2.2 &  2.8 \\
\ariv\   & 4711 & \dots             &   4.8$\pm$1.3  &    5.4$\pm$0.3 &    5.8$\pm$0.3 &    5.4 &     3.4 &    1.3 &    11.3 &     8.1 &  4.7 \\
\ariv\   & 4740 & \dots             & \dots             &    4.1$\pm$0.3 &    4.3$\pm$0.3 &    3.9 &     2.5 &    1.0 &     8.2 &     5.8 &  3.4 \\
\oiii\   & 4959 &  347$\pm$18    &   385$\pm$120  &     353$\pm$18 &    310$\pm$16  &    419 &     398 &    347 &     509 &     504 &  480\\
\oiii\   & 5007 & 1037$\pm$56    &  1169$\pm$62   &   1049$\pm$53  &    926$\pm$47  &   1250 &    1188 &   1034 &    1518 &    1503 &  1430\\
\cliii\  & 5518 &    1.5$\pm$0.9 & \dots             &    0.9$\pm$0.1 &    0.7$\pm$0.1 &    1.8 &     1.8 &    1.9 &     2.0 &     2.1 &  2.1 \\
\cliii\  & 5538 &    1.1$\pm$0.9 & \dots             &    0.6$\pm$0.1 &    0.6$\pm$0.1 &    1.5 &     1.5 &    1.5 &     1.6 &     1.6 &  1.6 \\
\nii\    & 5755 &    3.9$\pm$0.9 & \dots             &    0.2$\pm$0.1 &  \dots            &    0.2 &     0.3 &    0.7 &     0.2 &     0.2 &  0.4  \\
\hei\    & 5876 &    9.9$\pm$1.1 &   6.2$\pm$1.4  &    3.7$\pm$0.2 &    3.6$\pm$0.2 &    8.4 &     9.4 &   11.4 &     6.1 &     6.7 &  8.3 \\
\nii\    & 6548 &   60.6$\pm$3.1 &   7.1$\pm$1.2  &    3.4$\pm$0.2 &    2.7$\pm$0.2 &    3.8 &     4.2 &   10.7 &     1.9 &     2.1 &  4.8 \\
\ha\     & 6563 &    283$\pm$15  &   285$\pm$15   &    260$\pm$13  &    271$\pm$14  &    276 &     275 &    276 &     277 &     276 &  276 \\
\nii\    & 6584 &    187$\pm$10 &   21.9$\pm$1.5  &    9.8$\pm$0.5 &    7.8$\pm$0.4 &   11.2 &    12.6 &   31.4 &     5.6 &     6.3 &  14.1 \\
\hei\    & 6678 &    1.7$\pm$0.9 &   1.7$\pm$1.1  &    1.4$\pm$0.1 &    1.4$\pm$0.1 &    2.0 &     2.2 &    2.8 &     1.4 &     1.6 &  2.0 \\
\sii\    & 6716 &   17.4$\pm$1.2 &   5.4$\pm$1.1  &    2.7$\pm$0.2 &    2.2$\pm$0.1 &    2.9 &     3.3 &    6.3 &     2.1 &     2.4 &  4.4 \\
\sii\    & 6731 &   15.5$\pm$1.1 &   4.2$\pm$1.1  &    2.4$\pm$0.2 &    2.0$\pm$0.1 &    3.0 &     3.4 &    6.4 &     2.0 &     2.2 &  4.1 \\
\hei\    & 7065 &    2.4$\pm$1.6 &  \dots            &  \dots            &    0.5$\pm$0.2 &    2.0 &     2.1 &    2.7 &     1.4 &     1.5 &  1.9 \\
\hline
\end{tabular}
\label{tab:emmision_lines}
\vspace{0.15cm}
\end{center}
\end{table*} 

Taking advantage of our high-quality optical spectra of NGC\,6905, which were used to study the physical properties and abundances of this PNe, in addition to the available IR and radio observations, we can test the predictions from our stellar atmosphere PoWR model using the photoionization code {\scshape Cloudy} \citep[version 17.01;][]{Ferland2017}. {\scshape Cloudy} calculates the emissivity of the gas taking into account the dust present in the nebula. Furthermore, using the {\scshape pyCloudy} routines \citep{Morisset2013} we can produce synthetic optical and IR observations from pseudo 3D models which can be directly compared to the observations presented here. 

In order to create a 3D model of NGC\,6905, ten 1D models were generated by varying the external radius of each of them as a function of the latitudinal angle $\theta$ with respect to the equatorial plane (from 0$^{\circ}$ to 90$^{\circ}$ with steps of 10 degrees). Subsequently, {\sc pyCloudy} re-computes the 1D models and builds the pseudo 3D model. A detailed description of this process can be found in \cite{Gesicki2016}.

{\scshape Cloudy} requires as an input: the source of ionization (spectral shape and $L_{\star}$), the density distribution of the nebula (geometry, density and chemical abundances) and the dust properties (size distribution and composition). For this, we used the NLTE stellar atmosphere PoWR model of HD\,193949 discussed in Section~3.2 as the ionizing source. The morphology has been defined as an ellipsoidal shape with a spherical cavity as suggested by the NOT and {\itshape Spitzer} IRAC images \citep[see also][]{Phillips2010}. A semi-minor axis of 7.5$\times$10$^{17}$~cm and a semi-major axis of 1.5$\times$10$^{18}$~cm. Two cases for the density have been selected, Model A with $n_\mathrm{e}$=600~cm$^{-3}$ and Model~B with $n_\mathrm{e}$=400~cm$^{-3}$, which are the limiting density values of the $n_\mathrm{e}$([S\,{\sc ii}]) listed in Table~\ref{tab:sample2}. Figure~\ref{fig:pycloudy_model} shows the density distribution of the {\scshape pyCloudy} 3D model. The rectangular regions in this figure labeled as S1, S2 and S3 show the relative position and apertures of the synthetic pseudo-slits which are representative of regions A1, A2, A3 and A4 defined in Figure~\ref{fig:ngc6909_opt}.

\begin{figure}
\begin{center}
\includegraphics[width=1.0\linewidth]{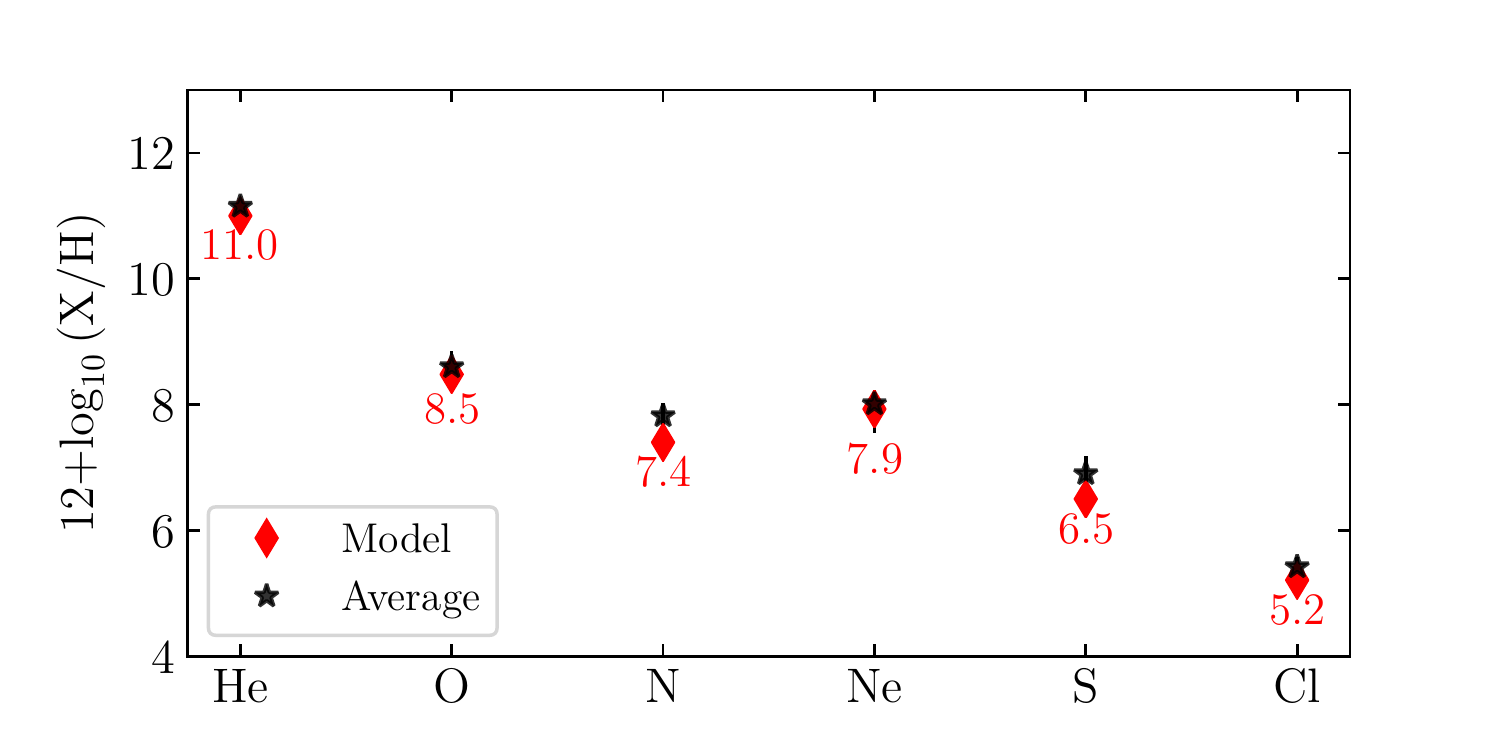}
\caption{Elemental abundances of NGC\,6905. Red diamonds shows the values used as {\sc Cloudy} input while black star is the averaged value in all slit positions. The length of the black vertical lines correspond to the minimum and maximum value observed in the nebula.}
\label{fig:abundances}
\end{center}
\end{figure}

To achieve a good fit to the emission lines, we had to tailor some of the elemental abundances. We started our model by adopting averaged abundance values obtained for A1--A4 and A6--A8 listed in the bottom rows of Table~\ref{tab:sample2}. However, some of these were varied in order to produce a good fit to the emission lines. Figure~\ref{fig:abundances} compares the abundances used in our best {\sc Cloudy} models in comparison with the averaged abundances from different elements. In all cases, the elemental abundances were slightly varied but still within the error bars of the averaged values.

\begin{figure}
\begin{center}
\includegraphics[width=1.0\linewidth]{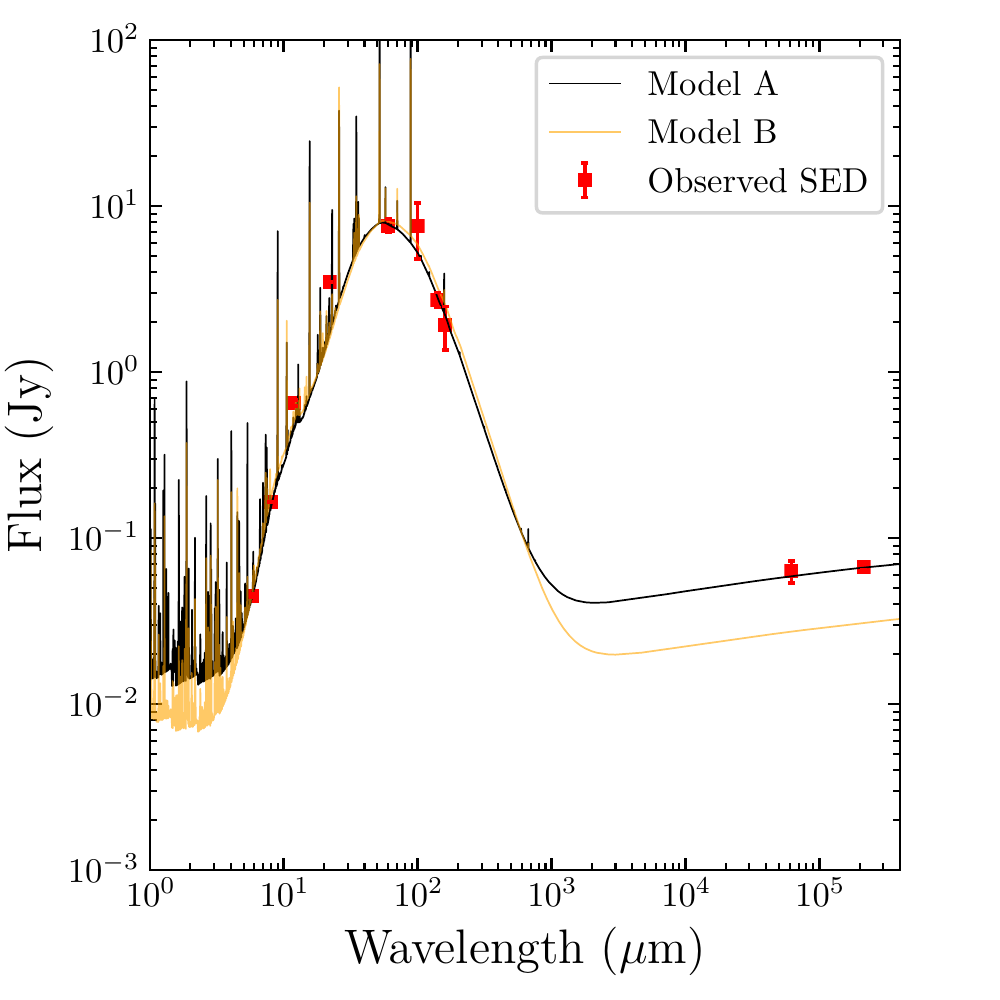}
\caption{Observed IR photometry of NGC\,6905 (red squares) compared to synthetic spectra obtained with {\sc Cloudy}. Model A fits the radio measurements whilst Model B appropriately describes H$\beta$ flux.}
\label{fig:sed_cont}
\end{center}
\end{figure}

Figure~\ref{fig:sed_cont} presents the observed SED listed in Table~\ref{tab:ir_photometry} which includes IR and radio measurements, this peaks at $\sim$70~$\mu$m. To model the SED of NGC\,6905, we included amorphous C in our {\scshape Cloudy} models with a power-law size distribution of $N(a)\propto a^{-3.5}$ \citep{Mathis1977}. Several combination of dust size distribution were attempted in order to produce a good fit to the SED. We found that it is necessary to include two populations of grains: a small population of dust with sizes $a_\mathrm{small}$=[0.001--0.002]~$\mu$m and a second one with $a_\mathrm{big}$=[0.01--0.05]~$\mu$m.

The peak of the IR emission is reproduced by the big grains while the small grains are necessary to fit the SED below 25~$\mu$m. In Figure~\ref{fig:sed_cont} we compare the synthetic photometry from our {\sc Cloudy} models to the observed SED. Both of our models fit most of the SED with a certain deviation for wavelengths between 15--30~$\mu$m. This problem was discussed by \citet{GLL2018} for the H-rich PN IC\,418 and by \citet{Toala2021} for the H-deficient born-again PNe A\,30 and A\,78. This lack of emission has been attributed to the presence of hydrogenated carbonaceous species not included in the current version of {\scshape Cloudy}. 

Model A is able to reproduce most of the optical emission lines as well as the SED down to radio frequencies. The emission line fluxes from our model are compared to those obtained from observations in Table~\ref{tab:emmision_lines}. Model A makes a good job fitting most of the emission lines, although we note that there is a large discrepancy with the \oiia\ emission line as our model exceeds 8 times the measured values from the central slits. The flux obtained from our model is more consistent with the outer regions of NGC\,6905 (slits A1 and A8). We speculate that the differences between our model and the observed emission lines might be attributed to shocks produced by the fast stellar wind of HD\,193949\footnote{Shock physics is not included in the current version of {\sc Cloudy}.}. The CSPN of WRPNe possess strong stellar winds and have bee reported to exhibit the highest expansion velocities compared to PN harboring H-rich CSPN \citep[see][]{Pena2003}.

The synthetic optical emission lines were used to estimate $T_\mathrm{e}$(\oiii), $T_\mathrm{e}$(\nii) and $n_\mathrm{e}$(\sii), which are also consistent with those estimated from observations (see top rows of Table~\ref{tab:model_results}). Finally, the estimated ionized mass of this model is 0.464~M$_{\odot}$ with a mass of dust of $\sim1.7\times10^{-3}$~M$_{\odot}$, for an averaged dust-to-gas ratio of 0.36 per cent by mass. A total mass of $M_\mathrm{TOT}$= $M_\mathrm{gas} + M_\mathrm{dust}$=0.466~M$_\odot$ results from this model. Details are listed in Table~\ref{tab:model_results}.

\begin{table}
\begin{center}
\setlength{\columnwidth}{0.08\columnwidth}
\setlength{\tabcolsep}{0.3\tabcolsep}
\caption{{\scshape Cloudy} model results.}
\begin{tabular}{lcccccc}
\hline
\multicolumn{1}{l}{Parameter} &
\multicolumn{3}{c}{Model A} &
\multicolumn{3}{c}{Model B} \\
\multicolumn{1}{l}{} &
\multicolumn{1}{c}{Slit 1} &
\multicolumn{1}{c}{Slit 2} &
\multicolumn{1}{c}{ Slit 3} &
\multicolumn{1}{c}{ Slit 1} &
\multicolumn{1}{c}{ Slit 2} &
\multicolumn{1}{c}{ Slit 3} \\ 
\hline
$T_\mathrm{e}$ (\oiii) [K]           &   12700   & 11970 &  11685  &   13300 &  12900 &  12400  \\
$T_\mathrm{e}$ (\nii) [K]            &   12300   & 11860 &  11734 &   12780 &  12700  &  12460 \\
$n_\mathrm{e}$ (\sii) [cm$^{-3}$]   & 510  & 487 &  486 &  720 &  710 &  700 \\
\multicolumn{1}{l}{log(H$\beta$) [erg~cm$^{-2}$~s$^{-1}$] } &
\multicolumn{1}{c}{ $-$12.53} &
\multicolumn{1}{c}{ $-$12.40} &
\multicolumn{1}{c}{ $-$13.00} &
\multicolumn{1}{c}{ $-$12.90} &
\multicolumn{1}{c}{ $-$12.76} &
\multicolumn{1}{c}{ $-$13.39} \\
\hline
\multicolumn{1}{l}{$M_{\mathrm{TOTAL}}$ [M$_{\odot}$]} &  
\multicolumn{3}{c}{4.66$\times10^{-1}$} &
\multicolumn{3}{c}{ 3.10$\times10^{-1}$} \\ 
\multicolumn{1}{l}{ $M_{\mathrm{Gas}}$ [M$_{\odot}$]} &  
\multicolumn{3}{c}{ 4.64$\times10^{-1}$} &
\multicolumn{3}{c}{ 3.07$\times10^{-1}$} \\ 
\multicolumn{1}{l}{$M_{\mathrm{Dust}}$ [M$_{\odot}$]} &  
\multicolumn{3}{c}{1.69$\times10^{-3}$} &
\multicolumn{3}{c}{ 2.24$\times10^{-3}$} \\ 
\hline
\multicolumn{1}{l}{$a_\mathrm{big}$ [M$_{\odot}$]} &
\multicolumn{3}{c}{1.46$\times10^{-3}$} &
\multicolumn{3}{c}{1.94$\times10^{-3}$} \\
\multicolumn{1}{l}{$a_\mathrm{small}$ [M$_{\odot}$]} &
\multicolumn{3}{c}{2.25$\times10^{-4}$} &
\multicolumn{3}{c}{2.98$\times10^{-4}$} \\
\hline
\end{tabular}
\label{tab:model_results}
\end{center}
\end{table}

One of the major difficulties of Model A was to simultaneously fit the H$\beta$ and the radio photometry. Both spectral features are directly related to the ionized mass of the model and there was a certain degeneracy between models. For this, Model~A was produced so that we can give priority to the radio measurements. This resulted in a H$\beta$ flux $\sim$2 times larger than the observed. For this, we used Model~B to try to fit the H$\beta$ flux. Model~B is also compared to the observed SED in Figure~\ref{fig:sed_cont} with a yellow solid line. The intensity of the synthetic emission lines of Model B are also listed in Table~\ref{tab:emmision_lines} and the global properties of the model are listed in Table~\ref{tab:model_results}.

The differences between Model A and B are evident in the SED. Model B cannot reproduce the radio measurements (see Fig.~\ref{fig:sed_cont}). However, the prediction for the H$\beta$ line flux are consistent with those extracted from regions A1--A4 (see Table~\ref{tab:sample}). Model A resulted in a $M_\mathrm{TOT}$ 50 per cent larger than that of Model B. Furthermore, although Model B produces slightly larger $T_\mathrm{e}$ than Model A, these are still consistent with values obtained from observations.

We conclude that a more accurate model might be achieved by further including clumps and filaments within NGC\,6905, but such an elegant model is out of the scope of the present paper.

\section{DISCUSSION}

The NOT ALFOSC spectrum of HD\,193949 presented here has an unprecedented quality compared to previous studies of this [WR]-type CSPN \citep[see][]{Pena1998,Acker2003}. We have been able to identify the three classic WR features (VB, BB and the RB) plus other 21 broad features also originating from the WR star. The presence of so many lines makes it complex to assign a spectral type under the currently accepted classification schemes \citep[e.g.,][]{Crowther1998,Acker2003}, but we note that the emission line ratios described in Section~3.1 suggests that HD\,193949 can be classified as a [WR]-type star of the oxygen sequence no later than [WO2]. 

In addition, we have detected the broad emission lines of \nevii\ at 4555~\AA\ and 5666~\AA, and \neviii\ at 6068~\AA. It has been demonstrated by \citet{Werner2007} that these features, initially attributed to
\ovii\ and \oviii, can be reproduced by model atmospheres containing \nevii\ and \neviii.
However, our stellar atmosphere modelling performed with the PoWR code confirms previous suggestions that HD\,193949 is a [WO]-type star with a temperature of $\sim$140~kK that could not provide such highly ionized O species. Furthermore, we identified the \neviii\ at 1163 and 1165~\AA\, in the {\itshape FUSE} observations presented in Fig.~\ref{fig:powr_uv}.

One cannot avoid to mention that the sub-classification schemes proposed for [WO]-type stars \citep{Crowther1998,Acker2003} are based on the alleged presence of the \ovii\ and \oviii\ lines. Considering this, the classification scheme needs to be updated for one that is based on high-resolution spectra that could include the comparison of more broad spectral lines including Ne lines of high ionization potential. Furthermore, this would also have an impact on the construction of new theoretical models for WRPNe.

The high signal-to-noise of the ALFOSC spectrum allowed us to extract spectra from different regions in NGC\,6905: two corresponding to the low-ionization knots located at the farthermost NW and SE regions (A1 and A8) of this PN plus 5 inner regions (A2, A3, A4, A6 and A7). As expected, the A1 and A8 spectra exhibit the presence of low-ionization species such as \nia\ and \oi\ lines. On the other hand, the high-ionization \arv\ ion is only detected in the innermost regions (A3, A4 and A6). The physical parameter analysis seems to suggest that NGC\,6905 has more or less constant $T_\mathrm{e}$ and $n_\mathrm{e}$ along the slit of the ALFOCS spectrograph.

\begin{figure}
\begin{center}
\includegraphics[width=\linewidth]{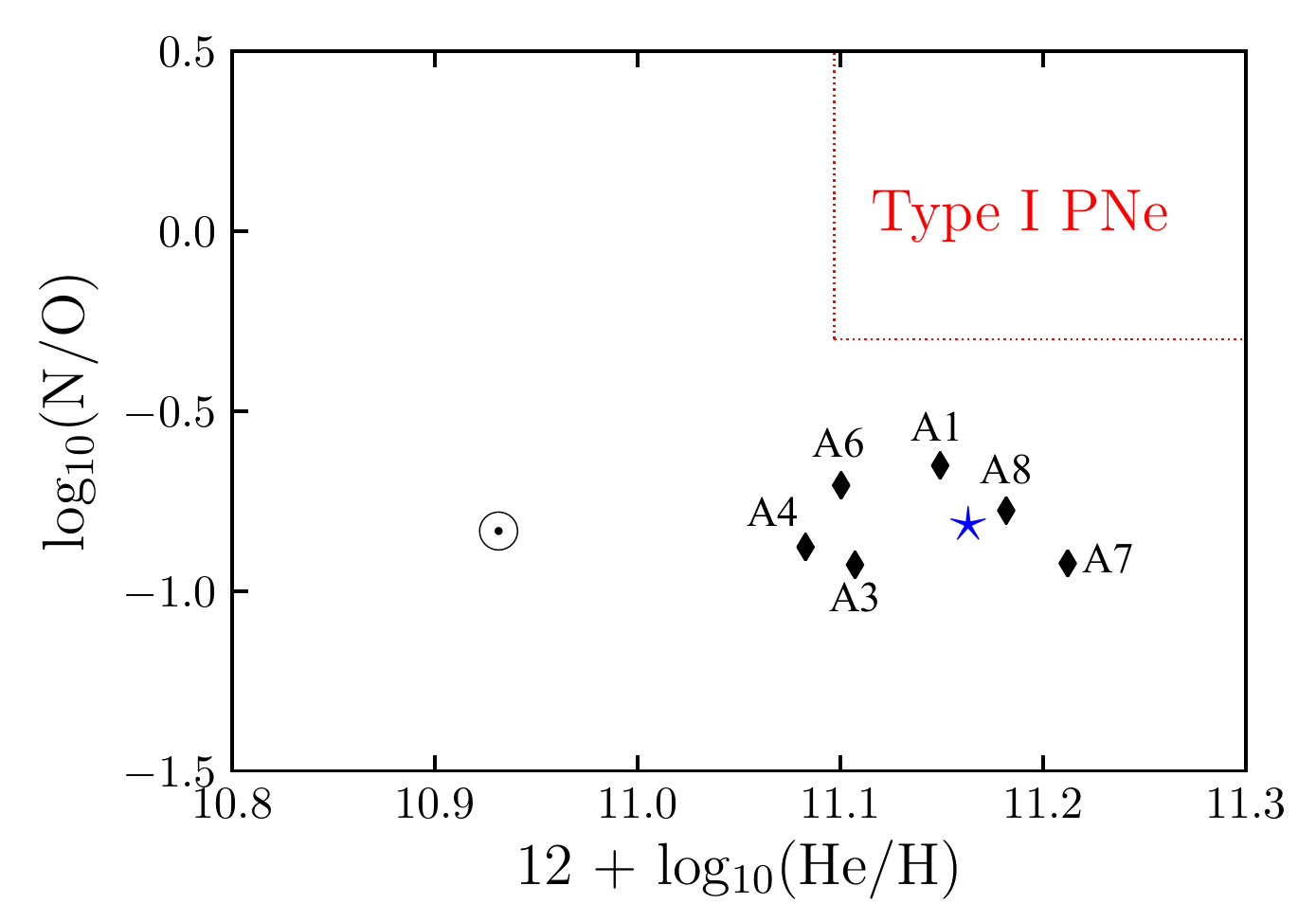}\\
\includegraphics[width=\linewidth]{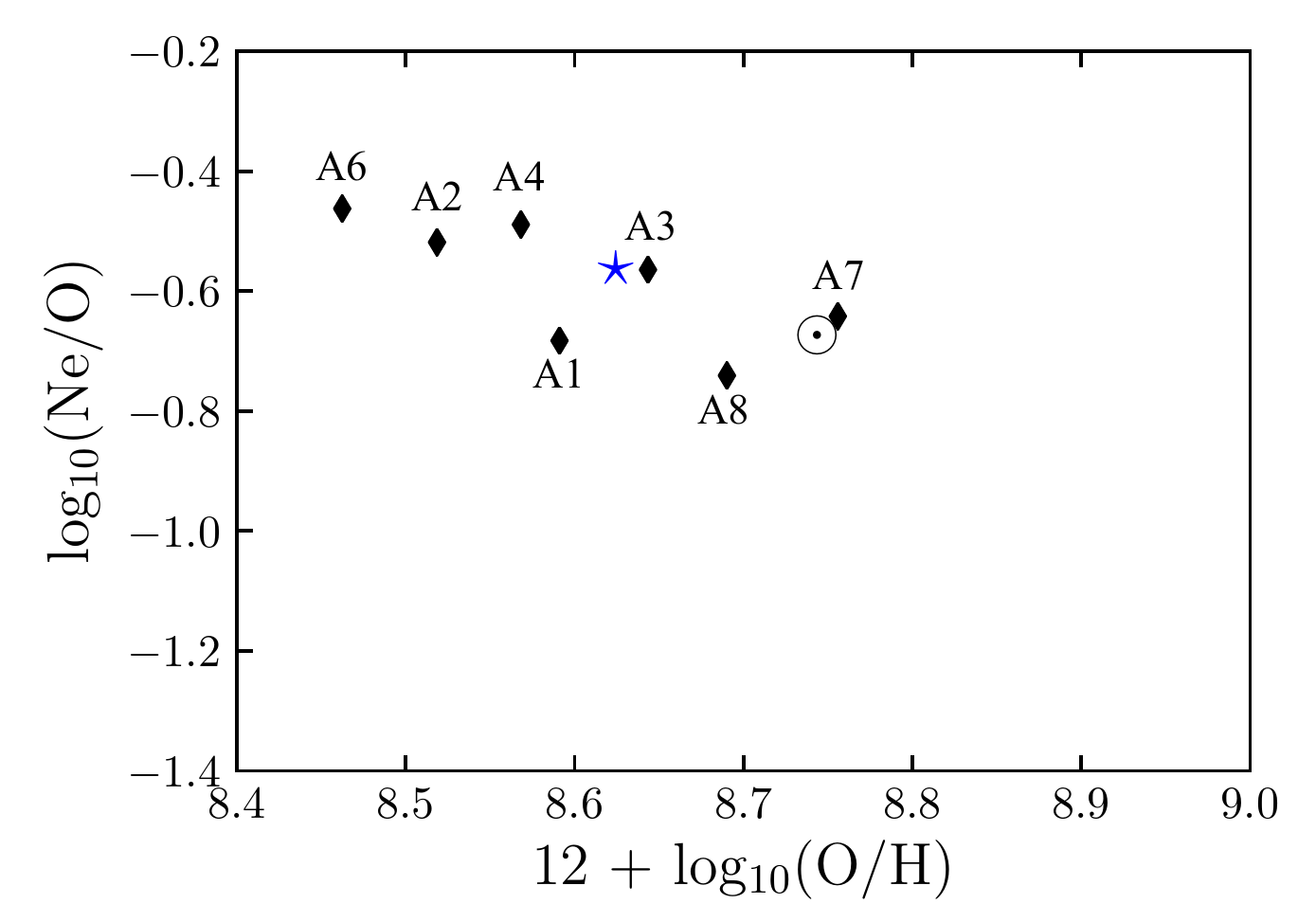}\\
\includegraphics[width=\linewidth]{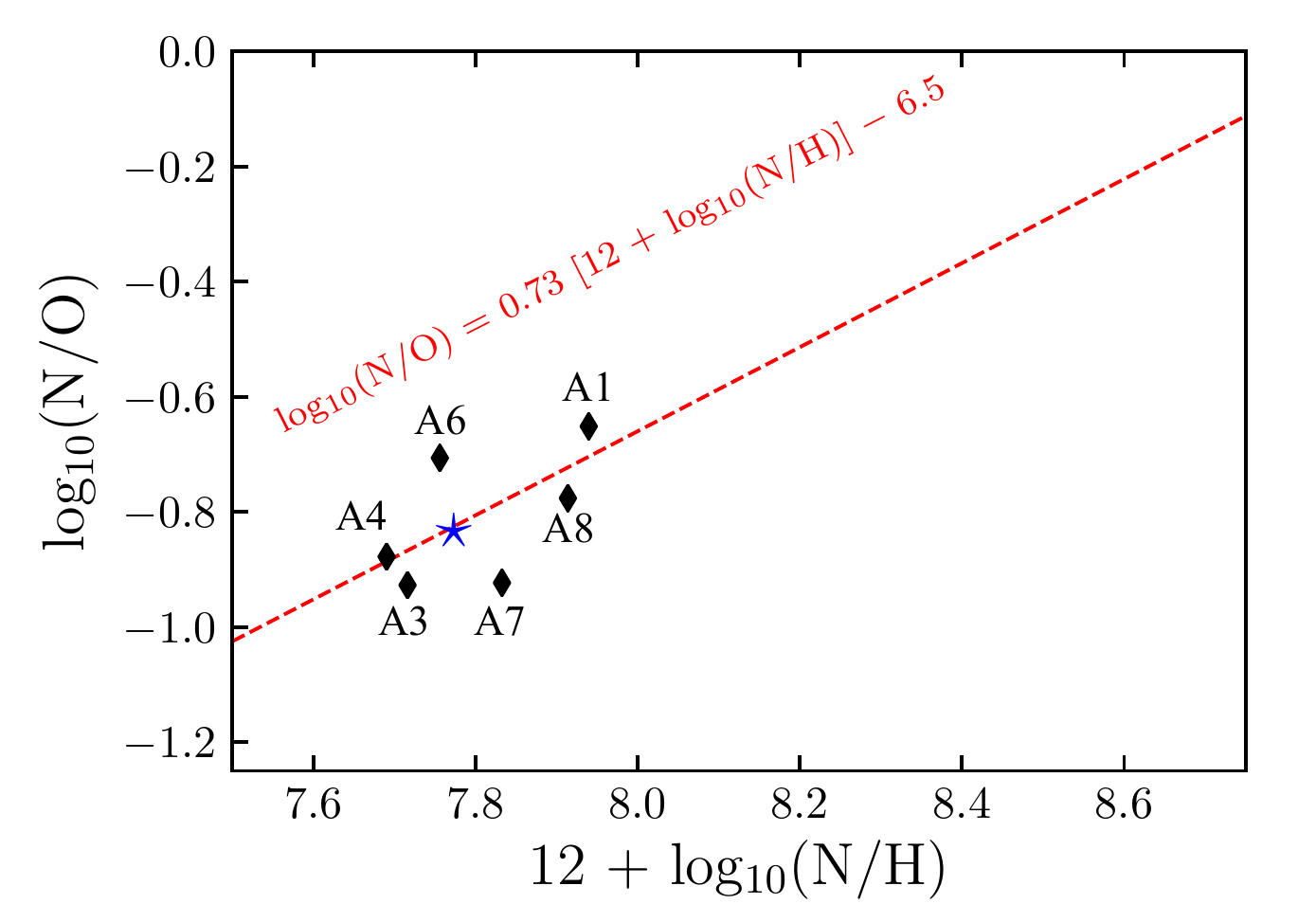}
\caption{Abundance ratio determinations of different elements obtained from values listed in Table~\ref{tab:sample2}. Diamonds represent values from slits A1--A8. The panels show the averaged value using all these slits in NGC\,6905 with a blue star. The solar values from \citet{Lodders2010} are also illustrated.
The top panel shows the N/O ratio versus the He abundance. Here the region for the Type~I PNe defined by \citet{Peimbert1990} is shown with red dotted-lines.
The middle panel displays the Ne/O ratio versus the O abundance.
The bottom panel presents the N/O ratio versus the N abundance.
The red dashed-line corresponds to the relation presented in \citet{GR2013} (see Eq.~(\ref{eq:GR2013}) and details in the text).
}
\label{fig:abund_figures}
\end{center}
\end{figure}

Furthermore, there is not a large difference between the total abundance determinations among the inner regions and the low-ionization knots. To facilitate a comparison, we have created the abundance ratio diagrams presented in Figure~\ref{fig:abund_figures}.
In the top panel we compare the N/O ratio versus the He abundance and in the middle panel we compare the Ne/O ratio versus the O abundance. Averaged values are shown with a blue star on each panel. None of the different regions within NGC\,6905 stands with different abundances. In fact, the averaged abundances are consistent with other WRPNe analysed in \citet{GR2013}. Fig.~\ref{fig:abund_figures} shows that our N/O values are located at the lower limits found by these authors, whilst the Ne/O ratio is within the reported values \citep[see figures~6 and 9 in][]{GR2013}.

According to the Peimbert classification \citep[][]{Peimbert1990,Peimbert2017}, Type~I PNe have ages $\sim$1~Gyr and initial masses $>$2~\msol, Type~II: $\sim$1--5~Gyr and $\sim$1.2--2~\msol, Type~III: $\sim$5~Gyr and $\sim$1.2~\msol\ and Type~IV: $>$5~Gyr and $<$1~\msol, in addition to being located in the Galactic halo.
Opposite to Type~II and III PNe, Type~I PNe present He- and N-rich envelopes (He/H$>$0.125 and N/O$>$0.5). Following \citet[][]{GR2013}, the region of the Type~I PNe is shown in Figure~\ref{fig:abund_figures} (top panel) as a reference. Our object is clearly far from that region. Just for completeness, the Galactic latitude of NGC\,6905 corresponds to values within the disk of the Galaxy ($l$, $b$ [deg] = 61.491253, $-$9.571280). Thus, a Type~IV classification is clearly ruled out.

We conclude that there is no anomalous C-enrichment within NGC\,6905 which suggests that no very late thermal pulse (VLTP) has been involved in the formation of this PNe or the production of its [WR] CSPN.
This is further corroborated by the very low N abundance of the stellar wind , $< 10^{-4}$ by mass, compared to what is expected from models accounting for the VLTP scenario of $\sim$0.1 by mass \citep[e.g.][]{Herwig2001,Althaus2005}.

Interestingly, the averaged values of 12$+$log$_{10}$(N/H)=7.75 and log$_{10}$(N/O)=$-$0.863 obtained for NGC\,6905 follow the relation described in \citet{GR2013}:
\begin{equation}
    \mathrm{log_{10}(N/O)} = 0.73 \left[ 12 + \mathrm{log_{10}(N/H)}\right]-6.5,
\label{eq:GR2013}
\end{equation}
\noindent with a difference of only $\approx$0.02~dex. We illustrate this in the bottom panel of Figure~\ref{fig:abund_figures}. \citet{GR2013} obtained Eq.~(\ref{eq:GR2013}) by fitting their observations. These authors argue that this relation follows from the fact that N enrichment occurs independently of the O abundance in PNe, implying that this is at the expense of C. Unfortunately, we are not able to estimate the C abundance to study this effect. \citet{GR2013} compare Eq.~(\ref{eq:GR2013}) with predictions from stellar evolution models presented by \citet{Karakas2010} in their figure~6 (lower panel). According to that figure, NGC\,6905 is placed at the lower left region of the diagram, suggesting that the progenitor mass of the CSPN of NGC\,6905 had an initial mass of $\sim$1~M$_{\odot}$.

With the synthetic SED obtained with our PoWR model we were able to calculate a {\scshape Cloudy} model for NGC\,6905 that reproduces quite well the observed nebular and dust properties.
We would like to remark that detailed models of the nebular and dust properties of PNe using a family of dust populations have been recently presented in the literature. In particular, we remark the work of \citet{GLL2018}. These authors included in their photoionization {\scshape Cloudy} models graphite, amorphous C, SiC and MgS to reproduce all the spectral features in the PN IC\,418. Such detailed analysis is out of the scope of this paper and only amorphous C is used in our {\scshape Cloudy} model which is then used to estimate the total mass of gas and dust. However, we note that including more dust species in the model, will not change dramatically the nebular properties and estimated masses.

Our {\scshape Cloudy} models suggest a total mass for NGC\,6905 in the range of 0.31~M$_{\odot}$ and 0.47~M$_{\odot}$ with a dust mass of $\approx2\times10^{-3}$~M$_\odot$, that is, a dust-to-gas ratio of 0.43--0.65 per cent. Adopting a current mass of 0.6~M$_{\odot}$ for HD\,193949 this amounts to an upper limit for the total mass of $\sim$1.07~M$_\odot$. Interestingly, our photoionization model is consistent with predictions from the N/O ratio versus N abundance. This suggests that most of the mass previously lost by the star is still in its vicinity and its currently photoionized by the strong UV flux from the CS. This seems to suggest that NGC\,6905 is one of the WRPN with the less massive progenitor star.

Finally, NGC\,6905 is included in the catalogue of new candidate binaries of \citet[][see their table A.1]{Chornay2021}, however, until there are comprehensive photometric follow-ups, the conclusive evidence of its probable binary nature is still lacking.
Recently \citet[][]{Jacoby2021} reported the identification of binary CSPNe with Kepler/K2 observations. NGC\,6905 is not included, however, given the richness of observations such as those of TESS, more studies regarding the binarity issue for this and other CSPNe are to be expected.

\section{SUMMARY AND CONCLUDING REMARKS}

We presented a multi-wavelength characterisation of the PN NGC\,6905 which harbours the [WR] CSPN HD\,193949. We combined UV, optical, IR and radio observations to fully characterise the physical properties of this WRPN NGC\,6905.
Our findings can be summarized as follows:

\begin{itemize}
\item The high-quality NOT ALFOSC spectrum of HD\,193949 allowed us to detect the three broad WR bumps, the so-called VB, BB and RB, confirming that this CSPN belongs to the [WO]-class of [WR] stars. Along with these classic broad features we also detected 21 WR features that include \heii, \civ, \ov\ and \ovi\ which suggest that the spectral type of HD\,193949 cannot be later than a [WO2]-subtype star.

\item Additionally to the two tens of WR features, we also detected broad emission features at
$\lambda$4555, $\lambda$5666 and $\lambda$6068~\AA, previously identified as emission lines from \ovii\ and \oviii, which according to more recent studies might be actually originated from stellar \nevii\ and \neviii. HD\,193949 is yet another CSPN with an effective temperature of $\sim$140~kK that could not provide such high ionization O species. We confirm the presence of the Ne lines in the {\itshape FUSE} spectrum of HD\,193949. We suggest that an updated classification scheme is needed, one that may also accounts for the Ne emission lines.

\item  We determined the main physical properties of the [WR] CSPN by using state-of-the-art non-LTE PoWR models. We analysed the optical spectrum and found that a model with $T_\mathrm{eff}$= $140$~kK
and surface-abundance pattern of H/He/C/N/O/Ne/Fe= $<$0.05/0.55/0.35/$<$0.000069/0.08/0.02/0.0014 (by mass) reproduces the observations sufficiently well. We confirm the high terminal velocity of the stellar wind of $\text{v}_{\infty}=2000\pm100~$km\,s$^{-1}$.

\item We studied the physical properties of different regions of NGC\,6905. We found that the low-ionization knots located at the NW and SE regions of NGC\,6905 do not exhibit different $n_\mathrm{e}$ nor $T_\mathrm{e}$ compared to the inner regions in this WRPN. We estimated averaged values of $n_\mathrm{e}$(\sii) = 500~cm$^{-3}$ with $T_\mathrm{e}$ ranging around 13~kK. 

\item The total abundances obtained from different regions within NGC\,6905 resulted in very similar values, which means that the low-density knots were not produced as part of a VLTP. NGC\,6905 has similar abundances as other WRPN, but slightly smaller N/O ratio. In particular, comparing the N/O ratio versus the N abundance following previous studies suggests that the CSPN of NGC\,6905 had a relatively low initial mass of $\sim$1~M$_\odot$. This makes NGC\,6905 one of the WRPN with the less massive central star.
We conclude that the underabundance of N in the stellar wind, the lack of 
C-enrichment in the PN and its chemically homogenous structure imply 
that NGC\,6905 is not the result of a VLTP.

\item Finally, we used {\scshape Cloudy} to produce a photoionization model of NGC\,6905 with the stellar atmosphere model obtained with PoWR for HD\,193949 as the ionization source and the abundances determined in this paper. Amorphous C is included in the calculations. Our model simultaneously reproduces the nebular and dust properties of NGC\,6905 and suggests that the total mass of gas of NGC\,6905 is in the range of 0.31~M$_{\odot}$ and 0.47~M$_{\odot}$ with a mass of dust of around $2.24\times10^{-3}$~M$_\odot$ and $1.69\times10^{-3}$~M$_\odot$ respectively. That is, a dust-to-gas ratio less than 0.7~per cent. Adopting a current mass of 0.6~M$_{\odot}$ for HD\,193949, we estimate that its initial mass was around 1.07~M$_{\odot}$, very similar to that estimated by comparing abundances with stellar evolution models.
\end{itemize}

\section*{Acknowledgements}

The authors thank the anonymous referee for a detailed report that helped clarify the presentation of the present work.
VMAGG acknowledges support from the Programa de Becas 
posdoctorales of the Direcci\'{o}n General 
de Asuntos del Personal Acad\'{e}mico (DGAPA) of the Universidad 
Nacional Aut\'{o}noma de M\'{e}xico (UNAM, Mexico). 
VMAGG and JAT acknowledge funding by DGAPA UNAM 
PAPIIT project IA100720. JAT also acknowledges support from the Marcos Moshinsky Fundation (Mexico). GR acknowledge support from Consejo Nacional de Ciencia y Tecnolog\'{i}a
(CONACyT) for student scholarship. MAG acknowledges support
of the Spanish Ministerio de Ciencia, Innovaci\'{o}n y Universidades 
grant PGC2018-102184-B-I00, cofunded by FEDER funds.
LS acknowledges funding by DGAPA UNAM  PAPIIT project IN-101819.
GR-L acknowledges support from CONACyT grant 263373 and PRODEP (Mexico).
This work is based on observations made with the Nordic Optical Telescope, 
operated by the Nordic Optical Telescope Scientific Association 
at the Observatorio del Roque de los Muchachos in La Palma, Spain, 
of the Instituto de Astrofisica de Canarias.
This work uses public data from the IR
telescope {\itshape Spitzer Space Telescope} through the NASA/IPAC Infrared
Science Archive, which is operated by the Jet Propulsion Laboratory
at the California Institute of Technology, under contract with the
National Aeronautics and Space Administration (NASA).
{\itshape WISE} is a joint project of the University of California (Los
Angeles, USA) and the JPL/Caltech.
The Infrared Astronomical Satellite ({\itshape IRAS}) was a joint project of the US, 
UK and the Netherlands.
This research is based on observations with {\itshape AKARI}, a JAXA 
project with the participation of ESA.
This work has make extensive use of the NASA's Astrophysics Data System.

\section*{Data availability}
The data underlying this work will be shared on reasonable request to the first author.


\end{document}